\documentclass[aps,prd,onecolumn,groupedaddress,longbibliography,nofootinbib]{revtex4-2}

\usepackage[T1]{fontenc}
\usepackage{amsmath}
\usepackage{amssymb}
\usepackage{verbatim}
\usepackage{mathrsfs}
\usepackage{braket}
\usepackage{mathtools}
\usepackage{acronym}
\usepackage[labelfont=bf]{caption}
\usepackage{subcaption}
\usepackage{textcomp}
\usepackage{tcolorbox}
\usepackage{tikz}
\usetikzlibrary{decorations.pathreplacing}
\makeatletter
\renewcommand{\l@section}{\@dottedtocline{1}{1.5em}{2.6em}}
\renewcommand{\l@subsection}{\@dottedtocline{2}{4.0em}{3.6em}}
\renewcommand{\l@subsubsection}{\@dottedtocline{3}{7.4em}{4.5em}}
\makeatother

\newcommand{\Reviewers}{0}
\newcommand{\Color}{}
\newcommand{\Reviewer}{}
\newcommand{\Review}[3]{
	\if\Reviewers1
	\renewcommand{\Reviewer}{#1}
	\if\Reviewer1
	\renewcommand{\Color}{green}
	\fi
	\if\Reviewer2
	\renewcommand{\Color}{cyan}
	\fi
	\if\Reviewer3
	\renewcommand{\Color}{magenta}
	\fi
	\if\Reviewer4
	\renewcommand{\Color}{orange}
	\fi
	\begin{tcolorbox}[colback=\Color,colframe=black]
		#2
		\vspace{8pt}
		\hrule
		\vspace{8pt}
		#3
	\end{tcolorbox}
	\fi
}

\newacro{CSB}[CSB]{Cauchy-Schwarz-Bunyakovsky}

\newcommand{\zbra}[1]{\prescript{}{z}{\bra{#1}}}
\newcommand{\zbraket}[1]{\prescript{}{z}{\braket{#1}}}

\newcommand{\parn}[3]{\frac{\partial^{#3} #1}{\partial #2^{#3}}}

\begin{document}


\title{On the Validity of the Weak Value Approximation}


\author{Benjamin No\"{e} Bauml}
\email[]{baumlb@oregonstate.edu}
\affiliation{Physics Department, Oregon State University, Corvallis, OR, 97331}


\date{\today}

\begin{abstract}
The weak value approximation has been in use for thirty-five years, but it has not as of yet received a truly complete derivation, leaving its mathematical validity in a state of limbo. Herein, I fill this gap, deriving the weak value approximation under the von Neumann and qubit probe models. Not only does this provide a level of mathematical support to the weak value approximation not attained in previous works, but the techniques demonstrated in the process might be usable by others to forge similar derivations for alternative models, thus teasing the possibility of even broader validation in the future.
\end{abstract}


\maketitle
\tableofcontents

\Review{1}{Comments from Reviewer 1 will be in these green boxes.}{Responses to Reviewer 1 will be below the line.}
\Review{2}{Comments from Reviewer 2 will be in these cyan boxes.}{Responses to Reviewer 2 will be below the line.}
\Review{3}{Comments from Reviewer 3 will be in these magenta boxes.}{Responses to Reviewer 3 will be below the line.}
\Review{4}{Comments from the Editor will be in these orange boxes.}{Responses to the Editor will be below the line.}

\section{Introduction: the Gamut of Ersatz Derivations}
\subsection{The Making of the Weak Measurement}
Forth from the susurrus of the innumerable scientific ideas born in 1988 came the concept of the weak value \cite{PhysRevLett.60.1351}, a strangely pre- and postcognitive actor on the quantum scene. Traditionally, measurement is taught in the strong regime, effected by projective measurements which read out specific, certain values for the observed quantity, and collapse the quantum state into the eigenspace of that measurement outcome. However, when the coupling between the probe system and the quantum system to be measured is weak, neither state is perturbed significantly enough to give a definitive readout (for the probe) or collapse into a particular eigenspace (for the measured system). Many identical repetitions of the weak measurement protocol are necessary to build up the statistics of the probe outcome to the point that usable information can be pulled out of the averages.

To measure a weak \textit{value}, one can introduce a further extension of this idea, wherein a second, strong measurement is taken. The weak value of the observable measured in the weak measurement is conditioned both on the initial state of the system (the \textit{preselection}) and the final state after the strong measurement (the \textit{postselection}). In this way, it is hoped by some that the weak value may shed light on the history of the measured system between the strong measurements, without causing much disturbance to the system.

Today, the weak value stands proud with a strained grin---fully capable of important work (such as its use in weak value amplification \cite{krafczyk2021enhanced}), yet often mired (and mired since early on \cite{Leggett1989HowComment,Peres1989Measurements,Aharonov1989AVReply}) in the existential crises of foundational theoretical controversy (drowning in the deluge of arguments both for \cite{PhysRevA.87.052104,PhysRevA.89.024102,PhysRevLett.111.240402,Reznik2023Photons} and against \cite{Sokolovski2016Weak,Griffiths_2016,Hance2023Weak,Bhati2022Capture} the use of weak values in determining the past of a particle, for example). Like a great many, the weak value finds itself vexed by its inner demons, and it is this foundational core---the validity of the weak value approximation itself---that I now address.

Perhaps ye be not familiar with this concept, but I urge ye not to lose heart. The form of the weak value approximation I shall derive is the convergence result
\begin{equation}
	e^{i\frac{\epsilon}{\hbar}\hat{A}\hat{D}}\ket{\psi}\ket{\pi} \to \sum_{\phi}\ket{\phi}\braket{\phi|\psi}Ne^{i\frac{\epsilon}{\hbar}A_{w}(\phi|\psi)\hat{D}}\ket{\pi},
	\label{eq:ConvRes}
\end{equation}

with respect to the coupling strength parameter, $\epsilon$. Specifically, what I mean to show is that the vector difference of the left and right hand sides of Eqn. \ref{eq:ConvRes} can be made arbitrarily small in the norm of the Hilbert space. A small (yet finite and nonzero) value of $\epsilon$ can be found such that the difference is at or below any desired magnitude.

Ultimately, I will seek to restrict this convergence result to the choice of a single postselection. However, since that restriction only differs from Eqn. \ref{eq:ConvRes} by a final projection onto a specific $\ket{\phi}$, proving the above form is sufficient to justify this restricted form (seen in Eqns. \ref{eq:NeumannSolved} and \ref{eq:QubitSolved}).

You may look at Eqn. \ref{eq:ConvRes} and the many esoteric symbols I have thrown your way with some apprehension, but fear not, for each piece shall be defined in time. 
The general point of this equation is as follows; on the left, the two systems will become entangled based on the eigenvalues of $\hat{A}$, while on the right, they are instead entangled based on $A_{w}(\phi|\psi)$, the weak value of $\hat{A}$. The validity of this approximation is the main inspiration for certain interpretations of the meaning of the weak value---interpretations I neither ascribe to, nor wish to focus on---therefore it is important to make sure this approximation is properly supported by mathematics, and to outline how it has been speciously presented in prior work.

To see the flaws of old and contemporary approaches, we must establish the mathematics from whence arose the original approximation. As one might expect, all things begin with a Hamiltonian, this one describing the interaction between two quantum systems. First, there is the \textit{system proper}, beginning in state $\ket{\psi}$, of which the experimenter wishes to measure the observable $\hat{A}$. Second, there is the \textit{probe}, beginning in state $\ket{\pi}$, whose observable $\hat{D}$ shall be measured to extract information about how $\ket{\pi}$ was disturbed in the interaction. One Hamiltonian which effects the desired interaction is
\begin{equation}
	\hat{H}(t) = -\epsilon\delta(t-t')\hat{A}\otimes\hat{D},
	\label{eq:NonHam}
\end{equation}
where $\epsilon$ is the parameter that controls the strength of the interaction (kept incredibly small for weak interactions), and $\delta(t-t')$ ensures that the interaction occurs precisely at some time $t'$, when it is desired. Hamiltonians of this and similar forms can be found in \cite{PhysRevLett.60.1351,Mello_2014,Bauml2022Much}. I chose to use a delta function, despite its unphysical nature, for the sake of mathematical simplicity, but any normalized function $g(t)$ which has compact support around the time of the interaction could serve, and the ultimate result would be the same. Though the details are fascinating (and \cite{Mello_2014} is happy to provide them, as was I in \cite{Bauml2022Much} to reproduce a measure of them), for now rest assured that the propagator effecting this interaction is
\begin{equation}
	\hat{U} = e^{i\frac{\epsilon}{\hbar}\hat{A}\hat{D}}.
\end{equation}
The above propagator assumes that $t > t'$ (so the interaction actually happens), and the $\otimes$ symbol has also been dropped to avoid clutter. Going forward, we will assume there is no other time evolution in the component systems that would evolve them independently (thus there would be nothing but the interaction term in the overall Hamiltonian). After the interaction takes place, our composite system---originally in the state $\ket{\psi}\ket{\pi}$---is transformed into
\begin{equation}
	e^{i\frac{\epsilon}{\hbar}\hat{A}\hat{D}}\ket{\psi}\ket{\pi}.
	\label{eq:Interact}
\end{equation}

So far, I have only preselected the state of the system proper, initializing it in the state $\ket{\psi}$. Now, the time has come to also postselect a final state. Suppose that, once the hullabaloo of our experimental apparatus is all said and done, we find from a projective measurement that the system proper has collapsed into the state $\ket{\phi}$. After this point (but without yet renormalizing), we find the composite system to be in the state
\begin{equation}
	\ket{\phi}\bra{\phi}e^{i\frac{\epsilon}{\hbar}\hat{A}\hat{D}}\ket{\psi}\ket{\pi}.
	\label{eq:PostInteract}
\end{equation}
This is the last step before the plunge into the approximation.

\subsection{The Expedited Birth of the Weak Value}
It always begins with a Taylor expansion of the propagator:
\begin{equation}
	\begin{split}
		\ket{\phi}\bra{\phi}e^{i\frac{\epsilon}{\hbar}\hat{A}\hat{D}}\ket{\psi}\ket{\pi} & = \ket{\phi}\bra{\phi}\sum_{j=0}^{\infty}\frac{1}{j!}\left(i\frac{\epsilon}{\hbar}\hat{A}\hat{D}\right)^{j}\ket{\psi}\ket{\pi} \\
		& = \ket{\phi}\sum_{j=0}^{\infty}\frac{1}{j!}\left(i\frac{\epsilon}{\hbar}\hat{D}\right)^{j}\braket{\phi|\hat{A}^{j}|\psi}\ket{\pi}.
	\end{split}
	\label{eq:Taylor}
\end{equation}
At this moment, the sum can be viewed as an operator solely on the state space of the probe, which will produce a change that we desire to detect in our quest to learn about the property $\hat{A}$. Note that we very nearly have the classic form of the weak value in the terms of our sum:
\begin{equation}
	A_{w}(\phi|\psi) = \frac{\braket{\phi|\hat{A}|\psi}}{\braket{\phi|\psi}}.
	\label{eq:WeakDef}
\end{equation}

Originally, the approach at this point was to apply the von Neumann measurement model \cite{Mello_2014,vonNeumann1932mathematische,vonNeumann2018mathematical}, in which the probe state $\ket{\pi}$ is a Gaussian in position space. It was then claimed by the authors of \cite{PhysRevLett.60.1351} (who presented nary a hint as to how the condition ought to be derived) that if the spread $\Delta$ in the probe state's canonical position variable $q$ (corresponding to the continuous eigenvalue spectrum of the probe observable $\hat{D}$, which they specified to be $\hat{q}$) satisfies $\Delta\ll\max_{n}|\braket{\phi|\psi}|/|\braket{\phi|\hat{A}^{n}|\psi}^{1/n}$, then in some vague manner, the higher order terms of the sum can be excised, leaving a linear equation that can then be approximated by an exponential:
\begin{equation}
	\ket{\phi}\bra{\phi}e^{i\frac{\epsilon}{\hbar}\hat{A}\hat{D}}\ket{\psi}\ket{\pi} \approx \ket{\phi}\left(\braket{\phi|\psi}+i\frac{\epsilon}{\hbar}\braket{\phi|\hat{A}|\psi}\hat{D}\right)\ket{\pi} = \ket{\phi}\braket{\phi|\psi}\left(1+i\frac{\epsilon}{\hbar}\frac{\braket{\phi|\hat{A}|\psi}}{\braket{\phi|\psi}}\hat{D}\right)\ket{\pi} \approx \ket{\phi}\braket{\phi|\psi}e^{i\frac{\epsilon}{\hbar}A_{w}(\phi|\psi)\hat{D}}\ket{\pi}.
\end{equation}

Later attempts \cite{PhysRevA.41.11,Duck1989Sense,Parks2010weak,Aharonov_1991} at justifying the approximation made the change a little less reliant on linear approximation, but instead on dropping purportedly insignificant terms. The composite system state from Eqn. \ref{eq:Taylor} can be rewritten as
\begin{align}
	\ket{\phi}\bra{\phi}e^{i\frac{\epsilon}{\hbar}\hat{A}\hat{D}}\ket{\psi}\ket{\pi} &
	\begin{aligned}[t]
		& = \ket{\phi}\braket{\phi|\psi}\sum_{j=0}^{\infty}\frac{1}{j!}\left(i\frac{\epsilon}{\hbar}\hat{D}\right)^{j}\frac{\braket{\phi|\hat{A}^{j}|\psi}}{\braket{\phi|\psi}}\ket{\pi} \\
		& = \ket{\phi}\braket{\phi|\psi}\sum_{j=0}^{\infty}\frac{1}{j!}\left(i\frac{\epsilon}{\hbar}\hat{D}\right)^{j}(A_{w})^{j}\ket{\pi} + \ket{\phi}\braket{\phi|\psi}\sum_{j=2}^{\infty}\frac{1}{j!}\left(i\frac{\epsilon}{\hbar}\hat{D}\right)^{j}\left[(\hat{A}^{j})_{w}-(A_{w})^{j}\right]\ket{\pi},
	\end{aligned}
	\label{eq:AjwAwjDiff}
\end{align}
(where $(\hat{A}^{j})_{w}=\frac{\braket{\phi|\hat{A}^{j}|\psi}}{\braket{\phi|\psi}}$) and if the second sum is quite small, then
\begin{equation}
	\ket{\phi}\bra{\phi}e^{i\frac{\epsilon}{\hbar}\hat{A}\hat{D}}\ket{\psi}\ket{\pi} \approx \ket{\phi}\braket{\phi|\psi}\sum_{j=0}^{\infty}\frac{1}{j!}\left(i\frac{\epsilon}{\hbar}\hat{D}\right)^{j}(A_{w})^{j}\ket{\pi} = \ket{\phi}\braket{\phi|\psi}e^{i\frac{\epsilon}{\hbar}A_{w}(\phi|\psi)\hat{D}}\ket{\pi}.
\end{equation}
This was the approach taken by \cite{PhysRevA.41.11}, wherein it was claimed that the condition for dropping these terms was
\begin{equation}
	(2\Delta)^{j}\frac{\Gamma(j/2)}{(j-2)!}|(\hat{A}^{j})_{w}-(A_{w})^{j}| \ll 1,
\end{equation}
for all $j \geq 2$. Reading this condition made me as mimsy as a borogove, as there was no talk of how these terms decay (without decay, though a billionth is much less than one, infinitely many billionths add to infinity). Furthermore, it was found to be erroneous by \cite{Duck1989Sense} (who were echoed by \cite{Parks2010weak}), though it was repeated in \cite{Aharonov_1991}. The idea of dropping these terms is not tommyrot, but it is a matter that requires more care---care I intend to take.

More recent treatments nearly revisit the linear approximation of \cite{PhysRevLett.60.1351}. In particular, they come in after the Taylor expansion step and, instead of distributing to eventually replace $\hat{A}$ with $(\hat{A}^{j})_{w}$, directly approximate the operator sum:
\begin{equation}
	\sum_{j=0}^{\infty}\frac{1}{j!}\left(i\frac{\epsilon}{\hbar}\hat{A}\hat{D}\right)^{j} \approx 1 + i\frac{\epsilon}{\hbar}\hat{A}\hat{D}.
\end{equation}
Those who use this method \cite{Hance2023Weak,Svensson2013Pedagogical} reason that making $\epsilon$ small allows the Taylor series to be truncated, then follow up with
\begin{equation}
	\bra{\phi}\sum_{j=0}^{\infty}\frac{1}{j!}\left(i\frac{\epsilon}{\hbar}\hat{A}\hat{D}\right)^{j}\ket{\psi} \approx \braket{\phi|\psi} + i\frac{\epsilon}{\hbar}\braket{\phi|\hat{A}|\psi}\hat{D} = \braket{\phi|\psi}\left(1+i\frac{\epsilon}{\hbar}A_{w}\hat{D}\right) \approx \braket{\phi|\psi}e^{i\frac{\epsilon}{\hbar}A_{w}\hat{D}}.
	\label{eq:ThroughSum}
\end{equation}
This presentation is subtly different from the earlier versions of the approximation, as where those were handling numbers built from operators and inner products, this is an argument about approximating operators directly. While it may be argued that all of the expressions in Eqn. \ref{eq:ThroughSum} end up in the same place (as the identity operator) when $\epsilon\to0$, it is more difficult to justify that they are close to each other for small, finite values of $\epsilon$. Considerations about the spectrum and the operator norm, if it exists, should be discussed---and so they shall be.

These represent what I consider to be the most prominent versions of justifications of the approximation. However, 
I would be remiss if I did not mention \cite{Vaidman2017WVBeyond,Vaidman2017WVControversy} which superficially resembled my goal in this work. Indeed, they do move the approximation step into the probe state itself, which is a critical step in my derivation. However, the main goal of \cite{Vaidman2017WVBeyond} (the mathematics of which is partially repeated in \cite{Vaidman2017WVControversy}) is distinctly concerned with evaluating the first leading order of difference between the postselected probe state and a probe state with shifted mean (and the details provided for certain steps in that process are beyond sparse; \cite{Dziewior2016Concept} provides more details, particularly in Section 5.2.1, but sidesteps the mystifying Eqn. 8 found in \cite{Vaidman2017WVBeyond}). My work is specifically fixated upon validating the approximation for finitely small values of $\epsilon$, and on providing as many elucidating details as possible.

I also ought to recognize \cite{Shikano2012Measurements,Dziewior2019Universality,Kofman2011,Haapasalo2011Weak}, which bear some similarity to my stated goal. Three of them are focused not on the convergence of the state of the composite system, as I am, but on the expectation value of observables related to the probe pointer. Of those three, \cite{Shikano2012Measurements} gives the standard perturbative result on operators in their Eqn. (39) (which is the form I discussed in the lead up to my Eqn. \ref{eq:ThroughSum}), 
\cite{Dziewior2019Universality} abstracts the details of the probe state, 
and \cite{Kofman2011} gives the standard result in their Eqn. (2.35) without showing the steps of the approximation, 
then goes on to derive general distributions for the state of the composite system in Section 11 without reducing them to the approximate versions from the early literature.

The fourth and final paper \cite{Haapasalo2011Weak} deserves special mention, as it is distinct from the others in its intensely analytic and operator-theoretic approach to the topic. It is intriguing to bring in such tools to address this issue, and that is also what makes my work different from theirs. I remain firmly rooted in the standard notation of ground-level quantum mechanics, which assists in maintaining the accessibility of my work among my peers, and demonstrates that the problem can be addressed from this more baseline level.

\subsection{A Sense of Foreboding}
In Section \ref{sec:Derivaton}, I lay out my a derivation of the weak value approximation. In \ref{subsec:General}, I keep the mathematics as general as possible, progressing as far as I can without making significant assumptions about the sort of probe being used. I then lay out the two main types of probes to consider in \ref{subsec:Consider}: the von Neumann model (for which the approximation is derived in \ref{subsec:Neumann}) and the qubit model (for which the approximation is derived in \ref{subsec:Qubit}).

Neither of these derivations follow the path of Eqn. \ref{eq:AjwAwjDiff}, but for the sake of highlighting the correct way to carefully examine that difference, I have a limited alternative derivation in Appendix \ref{ap:Agnostic} (which is corrected from my work in \cite{Bauml2022Much}). Additional appendices provide supplementary calculations supporting the alternative derivation (Appendices \ref{ap:GammaWorld} and \ref{ap:SubExp}), plus a look at adjusting the qubit probe model to nearly attain one of the advantageous properties of the von Neumann model (Appendix \ref{ap:QubitSep}).

The main body of the paper ends with Section \ref{sec:Epilogue}, wherein I recapitulate the main points of Section \ref{sec:Derivaton} and tie up the narrative.

\section{A Derivation Gingerly Done \label{sec:Derivaton}}
\subsection{Preamble: With General Probe \label{subsec:General}}
By prescription of the existing literature \cite{PhysRevA.41.11}, one might think my goal would be to justify that there exists a condition (on $\epsilon$ in particular) such that the final state of the composite system (after the weak interaction and postselection) is approximately
\begin{equation}
	\ket{\phi}\braket{\phi|\psi}e^{i\frac{\epsilon}{\hbar}A_{w}(\phi|\psi)\hat{D}}\ket{\pi},
	\label{eq:OldFinal}
\end{equation}
which is presented without normalization for simplicity. However, this missing normalization is rather discomfiting, and also noteworthy, as  the lack of normalization is present in both an overt way, incarnate in the factor $\braket{\phi|\psi}$, and a subtle way, incarnate in the operator $e^{i\frac{\epsilon}{\hbar}A_{w}(\phi|\psi)\hat{D}}$. The overt way can be handled later with ease, so it is the subtle way I shall focus on now.

At first glance, the operator $e^{i\frac{\epsilon}{\hbar}A_{w}(\phi|\psi)\hat{D}}$ doesn't appear as though it should cause a normalization problem---is it not unitary after all? In fact, it is not, as $A_{w}(\phi|\psi)$ may in general be a complex number. The operator can be expressed in the form
\begin{equation}
	e^{i\frac{\epsilon}{\hbar}A_{w}(\phi|\psi)\hat{D}} = e^{i\frac{\epsilon}{\hbar}\left(\Re A_{w}(\phi|\psi)+i\Im A_{w}(\phi|\psi)\right)\hat{D}} = e^{i\frac{\epsilon}{\hbar}\Re A_{w}(\phi|\psi)\hat{D}}e^{-\frac{\epsilon}{\hbar}\Im A_{w}(\phi|\psi)\hat{D}},
\end{equation}
where $\Re A_{w}(\phi|\psi)$ and $\Im A_{w}(\phi|\psi)$ are the real and imaginary parts, respectively, of $A_{w}(\phi|\psi)$. While $e^{i\frac{\epsilon}{\hbar}\Re A_{w}(\phi|\psi)\hat{D}}$ remains unitary, $e^{-\frac{\epsilon}{\hbar}\Im A_{w}(\phi|\psi)\hat{D}}$ is not, and impacts the normalization of the probe state.

To ease our troubled minds, let us define a normalized probe state,
\begin{equation}
	\ket{\pi^{w}} \equiv Ne^{i\frac{\epsilon}{\hbar}A_{w}(\phi|\psi)\hat{D}}\ket{\pi} = Ne^{i\frac{\epsilon}{\hbar}\left(\Re A_{w}(\phi|\psi)+i\Im A_{w}(\phi|\psi)\right)\hat{D}}\ket{\pi},
	\label{eq:piw}
\end{equation}
Thus, it must be that
\begin{equation}
	N = \bra{\pi}e^{-2\frac{\epsilon}{\hbar}\Im A_{w}(\phi|\psi)\hat{D}}\ket{\pi}^{-\frac{1}{2}}.
	\label{eq:Normalize}
\end{equation}
One may note that this is an expectation value of the exponentiated operator conditioned on initial probe state $\ket{\pi}$, and since $e^{-2\frac{\epsilon}{\hbar}\Im A_{w}(\phi|\psi)x}$ is a convex function in $x$, we can apply Jensen's inequality \cite{Durrett2011Essentials} to obtain
\begin{equation}
	\bra{\pi}e^{-2\frac{\epsilon}{\hbar}\Im A_{w}(\phi|\psi)\hat{D}}\ket{\pi} = \mathbb{E}\left[\left.e^{-2\frac{\epsilon}{\hbar}\Im A_{w}(\phi|\psi)\hat{D}}\right|\pi\right] \geq e^{-2\frac{\epsilon}{\hbar}\Im A_{w}(\phi|\psi)\mathbb{E}\left[\left.\hat{D}\right|\pi\right]} = e^{-2\frac{\epsilon}{\hbar}\Im A_{w}(\phi|\psi)\braket{\pi|\hat{D}|\pi}}.
\end{equation}
If the expectation value of the probe observable $\hat{D}$ is finite, then this provides us with an upper bound on the normalization constant:
\begin{equation}
	N = \bra{\pi}e^{-2\frac{\epsilon}{\hbar}\Im A_{w}(\phi|\psi)\hat{D}}\ket{\pi}^{-\frac{1}{2}} \leq e^{\frac{\epsilon}{\hbar}\Im A_{w}(\phi|\psi)\braket{\pi|\hat{D}|\pi}}.
	\label{eq:NormBound}
\end{equation}
For small $\epsilon$, we can be assured that the normalization constant can exceed 1 by only a vanishingly small amount. However, this does not provide a lower bound, so our ability to control the exact size of the normalization constant is limited. Omitting it from calculations is not advisable.

With the subtle normalization problem resolved, I now return to the overt one. One might expect that I would simply introduce a new normalization constant to the state
\begin{equation}
	\ket{\phi}\braket{\phi|\psi}\ket{\pi^{w}},
	\label{eq:PostNewFinal}
\end{equation}
but it is actually more practical to stop short of projection onto a particular postselection $\ket{\phi}$. Instead, I will construct a superposition over all possible postselections. The set of all possible postselections (an orthonormal basis for the state space of the system proper) could in general be uncountably infinite, but in practice, due to limitations on resolution of measurement devices and the design of the experimenter, it is liable to be at most countably infinite in practice (and often finite, such as in the case of the nested Mach-Zehnder interferometer set up \cite{PhysRevA.87.052104,PhysRevLett.111.240402,PhysRevA.89.024102,Griffiths_2016,Bauml2022Much}). The desired superposition is achieved in the state
\begin{equation}
	\sum_{\phi}\ket{\phi}\braket{\phi|\psi}\ket{\pi^{w}},
	\label{eq:NewFinal}
\end{equation}
where it is important to remember that $\ket{\pi^{w}}$ depends on the particular postselection $\phi$, even though I don't indicate this explicitly for notational compactness. My goal is to prove that this is the approximate final state of the system just before postselection---that is, to prove that the exact state of Eqn. \ref{eq:Interact} is approximately equal to Eqn. \ref{eq:NewFinal} under some condition on $\epsilon$ (which was effectively the statement of Eqn. \ref{eq:ConvRes}). If we postselect for some specific $\ket{\phi}$ in both equations (without renormalizing), then Eqn. \ref{eq:Interact} becomes Eqn. \ref{eq:PostInteract}, and the orthonormality of the set of postselections will allow us to recover Eqn. \ref{eq:PostNewFinal} from Eqn. \ref{eq:NewFinal}, demonstrating that Eqns. \ref{eq:PostInteract} and \ref{eq:PostNewFinal} are approximately equal. That is the true endpoint of the weak value approximation.

I desire to sort the terms of Eqn. \ref{eq:NewFinal} based on the weak values of $\hat{A}$. To do so I shall let $\bar{w}>0$ be some constant, and split the sum into parts depending on how the magnitude of the weak value compares:
\begin{equation}
	\sum_{\phi}\ket{\phi}\braket{\phi|\psi}\ket{\pi^{w}} = \sum_{|A_{w}(\phi|\psi)|\leq \bar{w}}\ket{\phi}\braket{\phi|\psi}\ket{\pi^{w}} + \sum_{|A_{w}(\phi|\psi)|> \bar{w}}\ket{\phi}\braket{\phi|\psi}\ket{\pi^{w}}.
\end{equation}
I wish to do something similar with the non-approximate final state of the system before postselection, seen in Eqn. \ref{eq:Interact}. To achieve this, let the eigenvalues of $\hat{A}$ be denoted by $a$, and their associated eigenstates (which also form an orthonormal basis for the state space of the system proper) be $\ket{a}$. Here, we will ignore the possibility of degeneracy and use notation consistent with a discrete eigenvalue spectrum, though this latter assumption is unnecessary. We can insert the identity $\hat{I}=\sum_{a}\ket{a}\bra{a}$, then separate the sum based on how the magnitudes of the eigenvalues compare to some constant $\bar{a}>0$:
\begin{equation}
	e^{-i\frac{\epsilon}{\hbar}\hat{A}\hat{D}}\ket{\psi}\ket{\pi} = 
	\sum_{|a|\leq\bar{a}}\ket{a}\braket{a|\psi}\ket{\pi_{a}} + \sum_{|a|>\bar{a}}\ket{a}\braket{a|\psi}\ket{\pi_{a}},
	\label{eq:pia}
\end{equation}
where
\begin{equation}
	\ket{\pi_{a}} \equiv e^{-i\frac{\epsilon}{\hbar}a\hat{D}}\ket{\pi}.
\end{equation}

With these expressions, we can now quantitatively compare how closely the approximate state of Eqn. \ref{eq:NewFinal} compares with the exact state of Eqn. \ref{eq:Interact} by taking the norm of their difference, then splitting it into manageable pieces:
\begin{align}
	\left|\left|e^{-i\frac{\epsilon}{\hbar}\hat{A}\hat{D}}\ket{\psi}\ket{\pi}-\sum_{\phi}\ket{\phi}\braket{\phi|\psi}\ket{\pi^{w}}\right|\right| &
	\begin{aligned}[t]
		& = \left|\left|\sum_{a}\ket{a}\braket{a|\psi}\ket{\pi_{a}}-\sum_{\phi}\ket{\phi}\braket{\phi|\psi}\ket{\pi^{w}}\right|\right| \\
		& \leq \left|\left|\sum_{|a|\leq\bar{a}}\ket{a}\braket{a|\psi}\ket{\pi_{a}} - \sum_{|A_{w}(\phi|\psi)|\leq \bar{w}}\ket{\phi}\braket{\phi|\psi}\ket{\pi^{w}} \right|\right| \\
		& \quad + \left|\left|\sum_{|a|>\bar{a}}\ket{a}\braket{a|\psi}\ket{\pi_{a}}\right|\right| + \left|\left|\sum_{|A_{w}(\phi|\psi)|> \bar{w}}\ket{\phi}\braket{\phi|\psi}\ket{\pi^{w}}\right|\right|.
	\end{aligned}
	\label{eq:PrimeDiff}
\end{align}
The last two terms (which I will square for convenience), are the easiest to handle, giving
\begin{align}
	\left|\left|\sum_{|a|>\bar{a}}\ket{a}\braket{a|\psi}\ket{\pi_{a}}\right|\right|^{2} & = \sum_{|a|>\bar{a}}\braket{\psi|a}\braket{a|\psi}\braket{\pi_{a}|\pi_{a}} = \sum_{|a|>\bar{a}}|\braket{a|\psi}|^{2}, \label{eq:Asum0}\\
	\left|\left|\sum_{|A_{w}(\phi|\psi)|> \bar{w}}\ket{\phi}\braket{\phi|\psi}\ket{\pi^{w}}\right|\right|^{2} & = \sum_{|A_{w}(\phi|\psi)|> \bar{w}}\braket{\psi|\phi}\braket{\phi|\psi}\braket{\pi^{w}|\pi^{w}} = \sum_{|A_{w}(\phi|\psi)|> \bar{w}}|\braket{\phi|\psi}|^{2}. \label{eq:Wsum0}
\end{align}
Both $\sum_{a}|\braket{a|\psi}|^{2}$ and $\sum_{\phi}|\braket{\phi|\psi}|^{2}$ are sums of the squared amplitudes of all terms of $\ket{\psi}$ expressed in different bases, so they must equal 1.

Regarding how this pertains to Eqn. \ref{eq:Asum0}, we know that each eigenvalue $a$ must be finite, so for any term $|\braket{a|\psi}|^{2}$ there exists $\bar{a} \geq |a|$ which excludes that term from the sum. As such, in accordance with the normalization of $\ket{\psi}$, for all $\xi > 0$, there exists $\bar{a}>0$ such that $\sum_{|a|>\bar{a}}|\braket{a|\psi}|^{2}<\xi$. By extension, it also follows that $1-\sum_{|a|\leq\bar{a}}|\braket{a|\psi}|^{2}<\xi$ when this condition is satisfied.

As for Eqn. \ref{eq:Wsum0}, we must be a bit more careful, as weak values are not as simple as eigenvalues. Indeed, it is possible that $A_{w}(\phi|\psi)$, at least as a mathematical object, may be infinite, so the ways in which this occurs should be considered.

First, it may be possible to choose the set of postselections such that there is a state $\ket{\phi'}$ such that $\braket{\phi'|\psi}=0$. This is the most benign infinity that can manifest, as this term does not contribute to $\sum_{\phi}|\braket{\phi|\psi}|^{2}$. As such, its refusal to be excluded has no impact on the convergence of Eqn. \ref{eq:Wsum0}. This issue doesn't arise in single postselections, as it doesn't make physical sense to postselect an impossible outcome. Rather, this is a side-effect of my choice to sum over all possible postselections in an orthonormal family.

Second, it may be possible to choose the set of postselections such that there is a state $\ket{\phi'}$ such that $|\braket{\phi'|A|\psi}|$ is infinite. Fortunately, there are some circumstances which would prevent this from happening. According to the \ac{CSB} inequality, we can say that
\begin{equation}
	|\braket{\phi'|A|\psi}| \leq
	\begin{cases}
		||\ket{\phi'}||\ ||A\ket{\psi}||, \\
		||A\ket{\phi'}||\ ||\ket{\psi}||.
	\end{cases}
\end{equation}
Thus, if either $A\ket{\psi}$ or $A\ket{\phi}$ is normalizable, then $|\braket{\phi'|A|\psi}|<\infty$. In the case of a bounded operator \cite{Axler2020Measure}, we know that $||A\ket{\psi}|| \leq ||A||_{op}||\ket{\psi}|| < \infty$ (and the same for $\ket{\phi'}$), so finiteness is assured. Furthermore, $||A\ket{\psi}||^{2} = \braket{\psi|A^{2}|\psi}$ (and the same for $\ket{\phi'}$), so an observable whose square has finite expectation value (as would be the case with an observable with finite expectation value and standard deviation) with respect to the either preselection or postselection would have $|\braket{\phi'|A|\psi}|<\infty$. This is a reasonable situation to expect, and since we choose the set of postselections, we can guarantee this if we so choose (though not necessarily for all sets of postselections with regard to any observable). Additionally, it would be physically problematic to say the least if an observable---an operator that should map the Hilbert space to itself---turned a square-integrable function in the space into a non-square integrable function. As such, I will assume that $|\braket{\phi|A|\psi}|<\infty$ for all $\ket{\phi}$.

With this assumption in place, we now can assume that each $A_{w}(\phi|\psi)$ for which $|\braket{\phi|\psi}|^{2}\neq0$ is finite in magnitude, and therefore we can select $\bar{w}\geq|A_{w}(\phi|\psi)|$ which excludes the associated term $|\braket{\phi|\psi}|^{2}$ from the sum in Eqn. \ref{eq:Wsum0}. It follows that, for all $\xi>0$, there exists $\bar{w}>0$ such that $\sum_{|A_{w}(\phi|\psi)|> \bar{w}}|\braket{\phi|\psi}|^{2} < \xi$. By extension, it must be that $1-\sum_{|A_{w}(\phi|\psi)|\leq\bar{w}}|\braket{\phi|\psi}|^{2} < \xi$ when this condition is satisfied.

Now that I have established that Eqns. \ref{eq:Asum0} and \ref{eq:Wsum0} can be made negligibly small by my choice of $\bar{a}$ and $\bar{w}$, respectively, we can move on to understanding the last difference, which I square and expand below:
\begin{align}
	\left|\left|\sum_{|a|\leq\bar{a}}\ket{a}\braket{a|\psi}\ket{\pi_{a}} - \sum_{|A_{w}(\phi|\psi)|\leq \bar{w}}\ket{\phi}\braket{\phi|\psi}\ket{\pi^{w}} \right|\right|^{2} &
	\begin{aligned}[t]
		& = \sum_{|a|\leq\bar{a}}|\braket{a|\psi}|^{2} + \sum_{|A_{w}(\phi|\psi)|\leq \bar{w}}|\braket{\phi|\psi}|^{2} \\
		& - \sum_{|a|\leq\bar{a}} \sum_{|A_{w}(\phi|\psi)|\leq\bar{w}} \braket{\psi|a}\braket{a|\phi}\braket{\phi|\psi}\braket{\pi_{a}|\pi^{w}} \\
		& - \sum_{|a|\leq\bar{a}} \sum_{|A_{w}(\phi|\psi)|\leq\bar{w}} \braket{\psi|\phi}\braket{\phi|a}\braket{a|\psi}\braket{\pi^{w}|\pi_{a}}.
	\end{aligned}
	\label{eq:GeneralDiff}
\end{align}
I have already established that the first two terms are approximately equal to 1, so it is the last two terms---the double sums that are being subtracted---which must be understood, which requires understanding $\braket{\pi_{a}|\pi^{w}}$.

From Eqns. \ref{eq:piw}, \ref{eq:Normalize}, and \ref{eq:pia}, this inner product can be written as
\begin{equation}
	\braket{\pi_{a}|\pi^{w}} = \frac{\bra{\pi}e^{i\frac{\epsilon}{\hbar}\left(\Re A_{w}(\phi|\psi)-a+i\Im A_{w}(\phi|\psi)\right)\hat{D}}\ket{\pi}}{\sqrt{\bra{\pi}e^{-2\frac{\epsilon}{\hbar}\Im A_{w}(\phi|\psi)\hat{D}}\ket{\pi}}}.
	\label{eq:piapiw}
\end{equation}
As I did for $\hat{A}$, let the eigenvalues of $\hat{D}$ be denoted by $d$, and their associated eigenstates (which form an orthonormal basis in the state space of the probe) be $\ket{d}$. Inserting the identity $\hat{I}=\sum_{d}\ket{d}\bra{d}$ gives us
\begin{equation}
	\braket{\pi_{a}|\pi^{w}} = \frac{1}{\sqrt{\bra{\pi}e^{-2\frac{\epsilon}{\hbar}\Im A_{w}(\phi|\psi)\hat{D}}\ket{\pi}}}\sum_{d}|\braket{d|\pi}|^{2}e^{-\frac{\epsilon}{\hbar}\Im A_{w}(\phi|\psi)d}e^{i\frac{\epsilon}{\hbar}\left(\Re A_{w}(\phi|\psi)-a\right)d}.
\end{equation}
We already know that, for a detector whose expectation value for $\hat{D}$ is finite, the normalization (given by Eqn. \ref{eq:Normalize}) should approach 1, but the above sum is trickier to handle. We require more assumptions about the probe.

\subsection{Probe Models to Consider \label{subsec:Consider}}

There are two main probe models that I will consider. The early literature \cite{PhysRevLett.60.1351,PhysRevA.41.11,Aharonov_1991} and some work following \cite{Hance2023Weak,Svensson2013Pedagogical,Bauml2022Much} focuses on the von Neumann model, which I shall cover in Section \ref{subsec:Neumann}. Other papers \cite{Hofmann2021direct,Griffiths_2016,Svensson2013Pedagogical,Bauml2022Much} examine a mathematically simpler model wherein the probe system is a qubit, or two-state quantum system. If one knows the exact model, one can calculate the quantity $\braket{\pi_{a}|\pi^{w}}$ in terms of the weak value, then proceed from there with the bounding procedure. However, it is not as straightforward when one only gives a general description of the qualities of the probe.

As an example, we could suppose that $\hat{D}$ has a bounded eigenvalue spectrum---that there exists $\bar{d}>0$ such that $\bar{d}\geq |d|$ for all $d$ in the spectrum of $\hat{D}$. In the most general case, the eigenvalues could still lie on a continuum, but this does encompass the case where the probe system has finitely many states (corresponding to at most as many eigenvalues). Even if the eigenvalues are discrete, even if finite in number, this leaves open the possibility of degeneracy---even infinite degeneracy---in their associated eigenspaces. While decomposing $\hat{D}$ into a sum of projectors is tempting, since it employs strategies similar to those used previously in bounding $|a|$ by $\bar{a}$ and $|A_{w}(\phi|\psi)|$ by $\bar{w}$, it doesn't pan out effectively.

Let me demonstrate. Given that $\bar{d}$ exists (and recalling that $|A_{w}(\phi|\psi)|\leq\bar{w}$, as well as Eqn. \ref{eq:NormBound}), we can conclude that
\begin{align}
	|\braket{\pi_{a}|\pi^{w}}| &
	\begin{aligned}[t]
		& \leq \frac{1}{\sqrt{\bra{\pi}e^{-2\frac{\epsilon}{\hbar}\Im A_{w}(\phi|\psi)\hat{D}}\ket{\pi}}}\sum_{d}|\braket{d|\pi}|^{2}e^{-\frac{\epsilon}{\hbar}\Im A_{w}(\phi|\psi)d} \left|e^{i\frac{\epsilon}{\hbar}\left(\Re A_{w}(\phi|\psi)-a\right)d}\right| \\
		& = \frac{1}{\sqrt{\bra{\pi}e^{-2\frac{\epsilon}{\hbar}\Im A_{w}(\phi|\psi)\hat{D}}\ket{\pi}}}\sum_{d}|\braket{d|\pi}|^{2}e^{-\frac{\epsilon}{\hbar}\Im A_{w}(\phi|\psi)d} \\
		& \leq \frac{e^{\frac{\epsilon}{\hbar}\bar{w}\bar{d}}}{\sqrt{\bra{\pi}e^{-2\frac{\epsilon}{\hbar}\Im A_{w}(\phi|\psi)\hat{D}}\ket{\pi}}}\sum_{d}|\braket{d|\pi}|^{2} \\
		& = \frac{e^{\frac{\epsilon}{\hbar}\bar{w}\bar{d}}}{\sqrt{\bra{\pi}e^{-2\frac{\epsilon}{\hbar}\Im A_{w}(\phi|\psi)\hat{D}}\ket{\pi}}} \\
		& \leq e^{\frac{\epsilon}{\hbar}\bar{w}\bar{d}}e^{\frac{\epsilon}{\hbar}\Im A_{w}(\phi|\psi)\braket{\pi|\hat{D}|\pi}} \\
		& \leq e^{\frac{\epsilon}{\hbar}\bar{w}\left(\bar{d}+|\braket{\pi|\hat{D}|\pi}|\right)}.
	\end{aligned}
	\label{eq:awBound}
\end{align}
This quantity does not help us control $|\braket{\pi_{a}|\pi^{w}}|$, as \ac{CSB} already indicates that it must be less than 1, and this quantity is greater than 1.

\subsection{A von Neumann Probe \label{subsec:Neumann}}
A von Neumann probe \cite{vonNeumann1932mathematische,vonNeumann2018mathematical,Mello_2014} has a Gaussian initial state, which I shall express as
\begin{equation}
	\ket{\mathscr{Q}} = \int_{-\infty}^{\infty}dq\ \frac{1}{\sqrt[4]{2\pi\Delta^{2}}}e^{-\frac{q^{2}}{4\Delta^{2}}}\ket{q} = \int_{-\infty}^{\infty}dp\ \frac{1}{\sqrt[4]{2\pi(\hbar/2\Delta)^{2}}}e^{-\frac{p^{2}}{4(\hbar/2\Delta)^{2}}}\ket{p},
\end{equation}
where $q$ is the \textit{probe canonical position}, $p$ is the \textit{probe canonical momentum}, and $\Delta$ is a parameter controlling the spread of the Gaussian state. A small $\Delta$ means a sharply peaked position Gaussian and a wide momentum Gaussian. The probe observable will be $\hat{q}$, which is the observable associated with the canonical position. The states $\ket{q}$ are its eigenstates (such that $\hat{q}\ket{q} = q\ket{q}$).

For compactness of notation, it will be useful to define real numbers $\alpha$ and $\beta$ such that
\begin{equation}
	\frac{\epsilon}{\hbar}A_{w}(\phi|\psi) \equiv \alpha+i\beta.
	\label{eq:AlphaBeta}
\end{equation}
To move between position and momentum representations, it is helpful to recall that the inner product of a position state and a momentum state (under Dirac normalization \cite{McIntyre2012Quantum}) is
\begin{equation}
\braket{q|p}=\frac{1}{\sqrt{2\pi\hbar}}e^{iqp/\hbar},
\end{equation}
while the inner product of two position states or two momentum states makes a Dirac delta function.

The normalization constant for this probe, as per Eqn. \ref{eq:Normalize}, is
\begin{align}
	N &
	\begin{aligned}[t]
		& = \bra{\mathscr{Q}}e^{-2\beta\hat{q}}\ket{\mathscr{Q}}^{-\frac{1}{2}} \\
		& = \left(\frac{1}{\sqrt{2\pi\Delta^{2}}}\int_{-\infty}^{\infty}dq\ e^{-\frac{q^{2}}{2\Delta^{2}}-2\beta q}\right)^{-\frac{1}{2}} \\
		& = \left(\frac{e^{2\Delta^{2}\beta^{2}}}{\sqrt{2\pi\Delta^{2}}}\int_{-\infty}^{\infty}dq\ e^{-\frac{(q+2\Delta^{2}\beta)^{2}}{2\Delta^{2}}}\right)^{-\frac{1}{2}} \\
		& = e^{-\Delta^{2}\beta^{2}}
	\end{aligned}
	\label{eq:NeumannNorm}
\end{align}
This indicates that, in the position basis, the probe state after the weak interaction (which had been called $\ket{\pi^{w}}$ in Eqn. \ref{eq:piw}) is
\begin{align}
	\ket{\mathscr{Q}^{w}} & \begin{aligned}[t]
		& \equiv e^{i(\alpha+i\beta)\hat{q}-\Delta^{2}\beta^{2}}\ket{\mathscr{Q}} \\
		& = e^{i(\alpha+i\beta)\hat{q}-\Delta^{2}\beta^{2}}\int_{-\infty}^{\infty}dq\ \frac{1}{\sqrt[4]{2\pi\Delta^{2}}}e^{-\frac{q^{2}}{4\Delta^{2}}}\ket{q} \\
		& = \frac{e^{-\Delta^{2}\beta^{2}}}{\sqrt[4]{2\pi\Delta^{2}}}\int_{-\infty}^{\infty}dq\ e^{-\frac{q^{2}}{4\Delta^{2}}+i\alpha q-\beta q}\ket{q} \\
		& = \frac{1}{\sqrt[4]{2\pi\Delta^{2}}}\int_{-\infty}^{\infty}dq\ e^{-\frac{1}{4\Delta^{2}}(q+2\Delta^{2}\beta)^{2}+i\alpha q}\ket{q}.
	\end{aligned}
	\label{eq:Qw}
\end{align}
If one were to strongly (i.e. projectively) measure the probe's position, the outcomes would be distributed by the probability density
\begin{equation}
	\mathscr{P}(q) = \braket{\mathscr{Q}^{w}|q}\braket{q|\mathscr{Q}^{w}} = \frac{1}{\sqrt{2\pi\Delta^{2}}}e^{-\frac{1}{2\Delta^{2}}(q+2\Delta^{2}\beta)^{2}} = \frac{1}{\sqrt{2\pi\Delta^{2}}}e^{-\frac{1}{2\Delta^{2}}(q+2\Delta^{2}\frac{\epsilon}{\hbar}\Im A_{w}(\phi|\psi))^{2}},
	\label{eq:Pq}
\end{equation}
which indicates that the probe's position Gaussian had its center shifted by an amount proportional solely to the imaginary component of the weak value of $\hat{A}$. Examining the momentum basis in the same manner obtains
\begin{align}
	\ket{\mathscr{Q}^{w}} & \begin{aligned}[t]
		& = e^{i(\alpha+i\beta)\hat{q}-\Delta^{2}\beta^{2}}\ket{\mathscr{Q}} \\
		& = e^{i(\alpha+i\beta)\hat{q}-\Delta^{2}\beta^{2}}\int_{-\infty}^{\infty}dq\ \frac{1}{\sqrt[4]{2\pi\Delta^{2}}}e^{-\frac{q^{2}}{4\Delta^{2}}}\ket{q} \\
		& = e^{-\Delta^{2}\beta^{2}}\frac{1}{\sqrt[4]{2\pi\Delta^{2}}}\int_{-\infty}^{\infty}dq\ e^{-\frac{q^{2}}{4\Delta^{2}}+i\alpha q-\beta q}\ket{q} \\
		& = e^{-\Delta^{2}\beta^{2}}\frac{1}{\sqrt[4]{2\pi\Delta^{2}}}\frac{1}{\sqrt{2\pi\hbar}}\int_{-\infty}^{\infty}dp\ \int_{-\infty}^{\infty}dq\ e^{-\frac{q^{2}}{4\Delta^{2}}+i\alpha q-\beta q-i\frac{pq}{\hbar}}\ket{p} \\
		& = e^{-\Delta^{2}\beta^{2}}\frac{1}{\sqrt[4]{2\pi\Delta^{2}}}\frac{1}{\sqrt{2\pi\hbar}}\int_{-\infty}^{\infty}dp\ e^{\Delta^{2}\left[\beta+i\left(\frac{p}{\hbar}-\alpha\right)\right]^{2}}\int_{-\infty}^{\infty}dq\ e^{-\frac{1}{4\Delta^{2}}\left\{q+2\Delta^{2}\left[\beta+i\left(\frac{p}{\hbar}-\alpha\right)\right]\right\}^{2}}\ket{p} \\
		& = e^{-\Delta^{2}\beta^{2}}\frac{1}{\sqrt[4]{2\pi\Delta^{2}}}\frac{1}{\sqrt{2\pi\hbar}}\int_{-\infty}^{\infty}dp\ e^{\Delta^{2}\left[\beta+i\left(\frac{p}{\hbar}-\alpha\right)\right]^{2}}\int_{-\infty+2\Delta^{2}\left[\beta+i\left(\frac{p}{\hbar}-\alpha\right)\right]}^{\infty+2\Delta^{2}\left[\beta+i\left(\frac{p}{\hbar}-\alpha\right)\right]}dq\ e^{-\frac{q^{2}}{4\Delta^{2}}}\ket{p} \\
		& = e^{-\Delta^{2}\beta^{2}}\frac{2\Delta\sqrt{\pi}}{\sqrt[4]{2\pi\Delta^{2}}}\frac{1}{\sqrt{2\pi\hbar}}\int_{-\infty}^{\infty}dp\ e^{\Delta^{2}\left[\beta+i\left(\frac{p}{\hbar}-\alpha\right)\right]^{2}}\ket{p} \\
		& = e^{-\Delta^{2}\beta^{2}}\frac{1}{\sqrt[4]{2\pi\Delta^{2}}\sqrt[4]{\hbar^{2}/2^{2}\Delta^{4}}}\int_{-\infty}^{\infty}dp\ e^{\Delta^{2}\left[\beta^{2}+2i\left(\frac{p}{\hbar}-\alpha\right)\beta-\left(\frac{p}{\hbar}-\alpha\right)^{2}\right]}\ket{p} \\
		& = \frac{e^{-2i\Delta^{2}\alpha\beta}}{\sqrt[4]{2\pi(\hbar/2\Delta)^{2}}} \int_{-\infty}^{\infty}dp\ e^{-\frac{1}{4(\hbar/2\Delta)^{2}}(p-\alpha\hbar)^{2}+2i\Delta^{2}\beta\frac{p}{\hbar}}\ket{p}.
	\end{aligned}
	\label{eq:BaseFourier}
\end{align}
Strongly measuring the probe's momentum would give outcomes distributed by the probability density
\begin{equation}
	\mathscr{P}(p) = \braket{\mathscr{Q}^{w}|p}\braket{p|\mathscr{Q}^{w}} = \frac{1}{\sqrt{2\pi(\hbar/2\Delta)^{2}}}e^{-\frac{1}{2(\hbar/2\Delta)^{2}}(p-\alpha\hbar)^{2}} = \frac{1}{\sqrt{2\pi(\hbar/2\Delta)^{2}}}e^{-\frac{1}{2(\hbar/2\Delta)^{2}}(p-\epsilon\Re A_{w}(\phi|\psi))^{2}},
	\label{eq:PqII}
\end{equation}
which indicates that the probe's momentum Gaussian had its center shifted by an amount proportional solely to the real component of the weak value of $\hat{A}$.

Recall that, in Eqn. \ref{eq:pia}, I defined $\ket{\pi_{a}}$. For the von Neumann model, I shall rename this $\ket{\mathcal{P}_{a}}$, which is defined to be
\begin{equation}
	\ket{\mathcal{P}_{a}} \equiv e^{i\frac{\epsilon}{\hbar}a\hat{q}}\ket{\mathscr{Q}} = \int_{-\infty}^{\infty}dp\ \frac{1}{\sqrt[4]{2\pi(\hbar/2\Delta)^{2}}}e^{-\frac{1}{4(\hbar/2\Delta)^{2}}(p-a\epsilon)^{2}}\ket{p}.
\end{equation}
It then follows that 
\begin{equation}
	e^{i\frac{\epsilon}{\hbar}\hat{A}\hat{q}}\ket{\psi}\ket{\mathscr{Q}} = \sum_{a}\ket{a}\braket{a|\psi}e^{i\frac{\epsilon}{\hbar}a\hat{q}}\ket{\mathscr{Q}} = \sum_{a}\ket{a}\braket{a|\psi}\ket{\mathcal{P}_{a}}.
	\label{eq:RepTriv}
\end{equation}
This also allows one to calculate
\begin{align}
	\braket{\mathcal{P}_{a}|\mathscr{Q}^{w}} & \begin{aligned}[t]
		& = \bra{\mathscr{Q}}e^{-i\frac{\epsilon}{\hbar}a\hat{q}}e^{i(\alpha+i\beta)\hat{q}-\Delta^{2}\beta^{2}}\ket{\mathscr{Q}} \\
		& = \frac{1}{\sqrt{2\pi\Delta^{2}}}\int_{-\infty}^{\infty}dq\ e^{-\frac{1}{4\Delta^{2}}q^{2}-i\frac{\epsilon}{\hbar}aq-\frac{1}{4\Delta^{2}}(q+2\Delta^{2}\beta)^{2}+i\alpha q} \\
		& = \frac{1}{\sqrt{2\pi\Delta^{2}}}\int_{-\infty}^{\infty}dq\ e^{-\frac{1}{2\Delta^{2}}q^{2}-i\frac{\epsilon}{\hbar}aq-\beta q+i\alpha q-\Delta^{2}\beta^{2}} \\
		& = \frac{1}{\sqrt{2\pi\Delta^{2}}}\int_{-\infty}^{\infty}dq\ e^{-\frac{1}{2\Delta^{2}}\left[q+\Delta^{2}\beta-i\Delta^{2}\left(\alpha-\frac{\epsilon}{\hbar}a\right)\right]^{2}+\frac{1}{2\Delta^{2}}\left[\Delta^{2}\beta-i\Delta^{2}\left(\alpha-\frac{\epsilon}{\hbar}a\right)\right]^{2}-\Delta^{2}\beta^{2}} \\
		& = e^{-\frac{1}{2}\Delta^{2}\beta^{2}-\frac{1}{2}\Delta^{2}\left(\alpha-\frac{\epsilon}{\hbar}a\right)^{2}-i\Delta^{2}\left(\alpha-\frac{\epsilon}{\hbar}a\right)\beta} \frac{1}{\sqrt{2\pi\Delta^{2}}}\int_{-\infty+\Delta^{2}\beta-i\Delta^{2}\left(\alpha-\frac{\epsilon}{\hbar}a\right)}^{\infty+\Delta^{2}\beta-i\Delta^{2}\left(\alpha-\frac{\epsilon}{\hbar}a\right)}dq\ e^{-\frac{q^{2}}{2\Delta^{2}}} \\
		& = e^{-\Delta^{2}\left\{\frac{1}{2}\left[\beta^{2}+\left(\alpha-\frac{\epsilon}{\hbar}a\right)^{2}\right]+i\left(\alpha-\frac{\epsilon}{\hbar}a\right)\beta\right\}} \\
		& = e^{-\frac{\epsilon^{2}\Delta^{2}}{\hbar^{2}}\left\{\frac{1}{2}\left[\left(\Im A_{w}(\phi|\psi)\right)^{2}+\left(\Re A_{w}(\phi|\psi)-a\right)^{2}\right]+i\left(\Re A_{w}(\phi|\psi)-a\right)\Im A_{w}(\phi|\psi)\right\}}.
	\end{aligned}
	\label{eq:PaQw}
\end{align}

Armed with this information, it is now possible to carry forward Eqn. \ref{eq:GeneralDiff}, which in the current notation, is
\begin{align}
	\left|\left|\sum_{|a|\leq\bar{a}}\ket{a}\braket{a|\psi}\ket{\mathcal{P}_{a}} - \sum_{|A_{w}(\phi|\psi)|\leq \bar{w}}\ket{\phi}\braket{\phi|\psi}\ket{\mathscr{Q}^{w}} \right|\right|^{2} &
	\begin{aligned}[t]
		& = \sum_{|a|\leq\bar{a}}|\braket{a|\psi}|^{2} + \sum_{|A_{w}(\phi|\psi)|\leq \bar{w}}|\braket{\phi|\psi}|^{2} \\
		& - \sum_{|a|\leq\bar{a}} \sum_{|A_{w}(\phi|\psi)|\leq\bar{w}} \braket{\psi|a}\braket{a|\phi}\braket{\phi|\psi}\braket{\mathcal{P}_{a}|\mathscr{Q}^{w}} \\
		& - \sum_{|a|\leq\bar{a}} \sum_{|A_{w}(\phi|\psi)|\leq\bar{w}} \braket{\psi|\phi}\braket{\phi|a}\braket{a|\psi}\braket{\mathscr{Q}^{w}|\mathcal{P}_{a}}.
	\end{aligned}
	\label{eq:NeumannDiff}
\end{align}
Since the last two terms can be rewritten as
\begin{align}
	\begin{aligned}[t]
		& -\sum_{|a|\leq\bar{a}} \sum_{|A_{w}(\phi|\psi)|\leq\bar{w}} \left(\braket{\psi|a}\braket{a|\phi}\braket{\phi|\psi}\braket{\mathcal{P}_{a}|\mathscr{Q}^{w}} + \braket{\psi|\phi}\braket{\phi|a}\braket{a|\psi}\braket{\mathscr{Q}^{w}|\mathcal{P}_{a}}\right) \\
		& = -2\sum_{|a|\leq\bar{a}} \sum_{|A_{w}(\phi|\psi)|\leq\bar{w}} \Re\left\{\braket{\psi|a}\braket{a|\phi}\braket{\phi|\psi}\braket{\mathcal{P}_{a}|\mathscr{Q}^{w}}\right\},
	\end{aligned}
	\label{eq:DoubSum}
\end{align}
and the summand can be expressed as
\begin{align}
	\begin{aligned}[t]
		& \Re\left\{\braket{\psi|a}\braket{a|\phi}\braket{\phi|\psi}\braket{\mathcal{P}_{a}|\mathscr{Q}^{w}}\right\} \\
		& = e^{-\frac{\epsilon^{2}\Delta^{2}}{\hbar^{2}}\left\{\frac{1}{2}\left[\left(\Im A_{w}(\phi|\psi)\right)^{2}+\left(\Re A_{w}(\phi|\psi)-a\right)^{2}\right]\right\}} \Re\left\{\braket{\psi|a}\braket{a|\phi}\braket{\phi|\psi}e^{-i\frac{\epsilon^{2}\Delta^{2}}{\hbar^{2}}\left(\Re A_{w}(\phi|\psi)-a\right)\Im A_{w}(\phi|\psi)}\right\} \\
		& = e^{-\frac{\epsilon^{2}\Delta^{2}}{\hbar^{2}}\left\{\frac{1}{2}\left[\left(\Im A_{w}(\phi|\psi)\right)^{2}+\left(\Re A_{w}(\phi|\psi)-a\right)^{2}\right]\right\}} \left[\cos\left(\frac{\epsilon^{2}\Delta^{2}}{\hbar^{2}}\left(\Re A_{w}(\phi|\psi)-a\right)\Im A_{w}(\phi|\psi)\right)\Re\braket{\psi|a}\braket{a|\phi}\braket{\phi|\psi}\right. \\
		& \phantom{= e^{-\frac{\epsilon^{2}\Delta^{2}}{\hbar^{2}}\left\{\frac{1}{2}\left[\left(\Im A_{w}(\phi|\psi)\right)^{2}+\left(\Re A_{w}(\phi|\psi)-a\right)^{2}\right]\right\}}} \left.+\sin\left(\frac{\epsilon^{2}\Delta^{2}}{\hbar^{2}}\left(\Re A_{w}(\phi|\psi)-a\right)\Im A_{w}(\phi|\psi)\right)\Im\braket{\psi|a}\braket{a|\phi}\braket{\phi|\psi}\right],
	\end{aligned}
	\label{eq:RPaQw}
\end{align}
Eqn. \ref{eq:NeumannDiff} becomes
\begin{align}
	\begin{aligned}[t]
		& \left|\left|\sum_{|a|\leq\bar{a}}\ket{a}\braket{a|\psi}\ket{\mathcal{P}_{a\epsilon}}-\sum_{|A_{w}(\phi|\psi)|\leq\bar{w}}\ket{\phi}\braket{\phi|\psi}e^{i(\alpha+i\beta)\hat{q}-\Delta^{2}\beta^{2}}\ket{\mathscr{Q}}\right|\right|^{2} \\
		& = \sum_{|a|\leq\bar{a}}|\braket{a|\psi}|^{2} + \sum_{|A_{w}(\phi|\psi)|\leq\bar{w}}|\braket{\phi|\psi}|^{2} \\
		& - 2\sum_{|a|\leq\bar{a}}\sum_{|A_{w}(\phi|\psi)|\leq\bar{w}} e^{-\frac{\epsilon^{2}\Delta^{2}}{\hbar^{2}}\left\{\frac{1}{2}\left[\left(\Im A_{w}(\phi|\psi)\right)^{2}+\left(\Re A_{w}(\phi|\psi)-a\right)^{2}\right]\right\}} \cos\left(\frac{\epsilon^{2}\Delta^{2}}{\hbar^{2}}\left(\Re A_{w}(\phi|\psi)-a\right)\Im A_{w}(\phi|\psi)\right)\Re\braket{\psi|a}\braket{a|\phi}\braket{\phi|\psi} \\
		& - 2\sum_{|a|\leq\bar{a}}\sum_{|A_{w}(\phi|\psi)|\leq\bar{w}} e^{-\frac{\epsilon^{2}\Delta^{2}}{\hbar^{2}}\left\{\frac{1}{2}\left[\left(\Im A_{w}(\phi|\psi)\right)^{2}+\left(\Re A_{w}(\phi|\psi)-a\right)^{2}\right]\right\}} \sin\left(\frac{\epsilon^{2}\Delta^{2}}{\hbar^{2}}\left(\Re A_{w}(\phi|\psi)-a\right)\Im A_{w}(\phi|\psi)\right)\Im\braket{\psi|a}\braket{a|\phi}\braket{\phi|\psi}.
	\end{aligned}
	\label{eq:44Reborn}
\end{align}
Due to the bounds on these sums and the finiteness of $\bar{a}$ and $\bar{w}$, the experimenter can control the coupling (via $\epsilon$)---or the width of the position Gaussian (via $\Delta$)---to establish the bounds
\begin{equation}
	-\sqrt{\frac{\pi}{2}} < -\frac{\epsilon\Delta}{\hbar}(\bar{w}+\bar{a}) < \frac{\epsilon\Delta}{\hbar}\left(\left\{
	\begin{matrix}
		\Re \\ \Im
	\end{matrix}
	\right\}A_{w}(\phi|\psi)-a\right) < \frac{\epsilon\Delta}{\hbar}(\bar{w}+\bar{a}) < \sqrt{\frac{\pi}{2}}
	\label{eq:BetweenPies}
\end{equation}
for all $a$ and $\ket{\phi}$ satisfying the bounds on the sums. In turn, this establishes the following bounds on functions:
\begin{align}
	0 < e^{-\frac{\epsilon^{2}\Delta^{2}}{\hbar^{2}}(\bar{w}+\bar{a})^{2}} & < e^{-\frac{\epsilon^{2}\Delta^{2}}{\hbar^{2}}\left\{\frac{1}{2}\left[\left(\Im A_{w}(\phi|\psi)\right)^{2}+\left(\Re A_{w}(\phi|\psi)-a\right)^{2}\right]\right\}} < 1, \label{eq:ExpBound}\\
	0 < \cos\left(\frac{\epsilon^{2}\Delta^{2}}{\hbar^{2}}(\bar{w}+\bar{a})^{2}\right) & < \cos\left(\frac{\epsilon^{2}\Delta^{2}}{\hbar^{2}}\left(\Re A_{w}(\phi|\psi)-a\right)\Im A_{w}(\phi|\psi)\right) < 1, \label{eq:CosBound}\\
	-1 < -\sin\left(\frac{\epsilon^{2}\Delta^{2}}{\hbar^{2}}(\bar{w}+\bar{a})^{2}\right) & < \sin\left(\frac{\epsilon^{2}\Delta^{2}}{\hbar^{2}}\left(\Re A_{w}(\phi|\psi)-a\right)\Im A_{w}(\phi|\psi)\right) < \sin\left(\frac{\epsilon^{2}\Delta^{2}}{\hbar^{2}}(\bar{w}+\bar{a})^{2}\right) < 1. \label{eq:SinBound}
\end{align}
Combining Eqn. \ref{eq:ExpBound} with Eqn. \ref{eq:CosBound} gives the bound
\begin{equation}
	0 < e^{-\frac{\epsilon^{2}\Delta^{2}}{\hbar^{2}}(\bar{w}+\bar{a})^{2}} \cos\left(\frac{\epsilon^{2}\Delta^{2}}{\hbar^{2}}(\bar{w}+\bar{a})^{2}\right) < e^{-\frac{\epsilon^{2}\Delta^{2}}{\hbar^{2}}\left\{\frac{1}{2}\left[\left(\Im A_{w}(\phi|\psi)\right)^{2}+\left(\Re A_{w}(\phi|\psi)-a\right)^{2}\right]\right\}} \cos\left(\frac{\epsilon^{2}\Delta^{2}}{\hbar^{2}}\left(\Re A_{w}(\phi|\psi)-a\right)\Im A_{w}(\phi|\psi)\right) < 1,
	\label{eq:ExpCosBound}
\end{equation}
and combining Eqn. \ref{eq:ExpBound} with Eqn. \ref{eq:SinBound} gives the bound
\begin{equation}
	-\sin\left(\frac{\epsilon^{2}\Delta^{2}}{\hbar^{2}}(\bar{w}+\bar{a})^{2}\right) < e^{-\frac{\epsilon^{2}\Delta^{2}}{\hbar^{2}}\left\{\frac{1}{2}\left[\left(\Im A_{w}(\phi|\psi)\right)^{2}+\left(\Re A_{w}(\phi|\psi)-a\right)^{2}\right]\right\}}\sin\left(\frac{\epsilon^{2}\Delta^{2}}{\hbar^{2}}\left(\Re A_{w}(\phi|\psi)-a\right)\Im A_{w}(\phi|\psi)\right) < \sin\left(\frac{\epsilon^{2}\Delta^{2}}{\hbar^{2}}(\bar{w}+\bar{a})^{2}\right).
	\label{eq:ExpSinBound}
\end{equation}
These bounds are valuable, as for all $\xi>0$, and for all $\bar{a}>0$ and $\bar{w}>0$, there exists $\epsilon>0$ such that 
\begin{align}
	1-e^{-\frac{\epsilon^{2}\Delta^{2}}{\hbar^{2}}(\bar{w}+\bar{a})^{2}} \cos\left(\frac{\epsilon^{2}\Delta^{2}}{\hbar^{2}}(\bar{w}+\bar{a})^{2}\right) & < \xi, \label{eq:XiExpCos}\\
	\sin\left(\frac{\epsilon^{2}\Delta^{2}}{\hbar^{2}}(\bar{w}+\bar{a})^{2}\right) & < \xi. \label{eq:XiSin}
\end{align}
In turn, these squeeze the formulae in Eqns. \ref{eq:ExpCosBound} and \ref{eq:ExpSinBound} closer to 1 and 0, respectively, for all applicable values of $a$ and all applicable postselections $\ket{\phi}$. Now, Eqn. \ref{eq:NeumannDiff} can be written as
\begin{align}
	\begin{aligned}[t]
		& \left|\left|\sum_{|a|\leq\bar{a}}\ket{a}\braket{a|\psi}\ket{\mathcal{P}_{a\epsilon}}-\sum_{|A_{w}(\phi|\psi)|\leq\bar{w}}\ket{\phi}\braket{\phi|\psi}e^{i(\alpha+i\beta)\hat{q}-\Delta^{2}\beta^{2}}\ket{\mathscr{Q}}\right|\right|^{2} \\
		& = \sum_{|a|\leq\bar{a}}|\braket{a|\psi}|^{2} + \sum_{|A_{w}(\phi|\psi)|\leq\bar{w}}|\braket{\phi|\psi}|^{2} - 2\Re\bra{\psi}\left[\sum_{|a|\leq\bar{a}}\ket{a}\bra{a}\sum_{|A_{w}(\phi|\psi)|\leq\bar{w}}\ket{\phi}\bra{\phi}\right]\ket{\psi} \\
		& + 2\sum_{|a|\leq\bar{a}}\sum_{|A_{w}(\phi|\psi)|\leq\bar{w}} \left(1-e^{-\frac{\epsilon^{2}\Delta^{2}}{\hbar^{2}}\left\{\frac{1}{2}\left[\left(\Im A_{w}(\phi|\psi)\right)^{2}+\left(\Re A_{w}(\phi|\psi)-a\right)^{2}\right]\right\}} \cos\left(\frac{\epsilon^{2}\Delta^{2}}{\hbar^{2}}\left(\Re A_{w}(\phi|\psi)-a\right)\Im A_{w}(\phi|\psi)\right)\right)\Re\braket{\psi|a}\braket{a|\phi}\braket{\phi|\psi} \\
		& - 2\sum_{|a|\leq\bar{a}}\sum_{|A_{w}(\phi|\psi)|\leq\bar{w}} e^{-\frac{\epsilon^{2}\Delta^{2}}{\hbar^{2}}\left\{\frac{1}{2}\left[\left(\Im A_{w}(\phi|\psi)\right)^{2}+\left(\Re A_{w}(\phi|\psi)-a\right)^{2}\right]\right\}} \sin\left(\frac{\epsilon^{2}\Delta^{2}}{\hbar^{2}}\left(\Re A_{w}(\phi|\psi)-a\right)\Im A_{w}(\phi|\psi)\right)\Im\braket{\psi|a}\braket{a|\phi}\braket{\phi|\psi} \\
		& < 1 + 1 - 2\Re\bra{\psi}\left[\sum_{|a|\leq\bar{a}}\ket{a}\bra{a}\sum_{|A_{w}(\phi|\psi)|\leq\bar{w}}\ket{\phi}\bra{\phi}\right]\ket{\psi} \\
		& + 2\xi\sum_{|a|\leq\bar{a}}\sum_{|A_{w}(\phi|\psi)|\leq\bar{w}} \Re\braket{\psi|a}\braket{a|\phi}\braket{\phi|\psi} + 2\xi\sum_{|a|\leq\bar{a}}\sum_{|A_{w}(\phi|\psi)|\leq\bar{w}}\Im\braket{\psi|a}\braket{a|\phi}\braket{\phi|\psi} \\
		& = 2 - 2\Re\bra{\psi}\left[\sum_{|a|\leq\bar{a}}\ket{a}\bra{a}\sum_{|A_{w}(\phi|\psi)|\leq\bar{w}}\ket{\phi}\bra{\phi}\right]\ket{\psi} \\
		& + 2\xi\Re\bra{\psi}\left[\sum_{|a|\leq\bar{a}}\ket{a}\bra{a}\sum_{|A_{w}(\phi|\psi)|\leq\bar{w}}\ket{\phi}\bra{\phi}\right]\ket{\psi} + 2\xi\Im\bra{\psi}\left[\sum_{|a|\leq\bar{a}}\ket{a}\bra{a}\sum_{|A_{w}(\phi|\psi)|\leq\bar{w}}\ket{\phi}\bra{\phi}\right]\ket{\psi}.
	\end{aligned}
	\label{eq:NeumannDiffProgressed}
\end{align}
In doing this, I have actually jumped the gun slightly. I should have added a few more conditions on $\bar{a}$ and $\bar{w}$ before setting $\epsilon$, but the current version of the formula helps to illustrate what those conditions must be.

Consider the inner product
\begin{equation}
	\bra{\psi}\left[\sum_{|a|\leq\bar{a}}\ket{a}\bra{a}\sum_{|A_{w}(\phi|\psi)|\leq\bar{w}}\ket{\phi}\bra{\phi}\right]\ket{\psi}.
	\label{eq:Inner}
\end{equation}
Intuitively, one might expect that this will approach 1 as $\bar{a}$ and $\bar{w}$ approach infinity. The trick is to actually demonstrate it. First, for convenience, define the (unnormalized) state
\begin{equation}
	\ket{\psi(\bar{w})} \equiv \sum_{|A_{w}(\phi|\psi)|\leq\bar{w}}\ket{\phi}\braket{\phi|\psi}.
\end{equation}
It has already been established (just before Eqn. \ref{eq:GeneralDiff}) that $\braket{\psi|\psi(\bar{w})}$ will approach 1 as $\bar{w}$ approaches infinity, so it is necessary to show that $\sum_{|a|\leq\bar{a}}\braket{\psi|a}\braket{a|\psi(\bar{w})}$ approaches $\braket{\psi|\psi(\bar{w})}$ as $\bar{a}$ approaches infinity. It is clear that
\begin{equation}
	\braket{\psi|\psi(\bar{w})} = \sum_{|a|\leq\bar{a}}\braket{\psi|a}\braket{a|\psi(\bar{w})} + \sum_{|a|>\bar{a}}\braket{\psi|a}\braket{a|\psi(\bar{w})},
\end{equation}
so it is equivalent to show that $\sum_{|a|>\bar{a}}\braket{\psi|a}\braket{a|\psi(\bar{w})}$ approaches zero. This occurs if and only if
\begin{equation}
	\lim_{\bar{a}\to\infty}\left|\sum_{|a|>\bar{a}}\braket{\psi|a}\braket{a|\psi(\bar{w})}\right| = 0.
	\label{eq:PsiAPsiw}
\end{equation}
Note that
\begin{equation}
	\lim_{\bar{a}\to\infty}\left|\sum_{|a|>\bar{a}}\braket{\psi|a}\braket{a|\psi(\bar{w})}\right| \leq \lim_{\bar{a}\to\infty}\sum_{|a|>\bar{a}}\left|\braket{\psi|a}\braket{a|\psi(\bar{w})}\right|,
\end{equation}
so if Eqn. \ref{eq:PsiAPsiw} is not true---if the limit is finite and nonzero---then the limit of $\sum_{|a|>\bar{a}}\left|\braket{\psi|a}\braket{a|\psi(\bar{w})}\right|$ must also be nonzero. If this expression, monotonically decreasing in $\bar{a}$, does not go to zero, then there must exist terms that remain in the sum for all values of $\bar{a}$. However, that would indicate the existence of infinite eigenvalues of $\hat{A}$, which would be a contradiction. As such, Eqn. \ref{eq:PsiAPsiw} must be true, and thus one can conclude that, for all $\xi > 0$, there exists $\bar{a}>0$ such that
\begin{equation}
	\left|\braket{\psi|\psi(\bar{w})} - \sum_{|a|\leq\bar{a}}\braket{\psi|a}\braket{a|\psi(\bar{w})}\right| < \frac{\xi}{2}.
\end{equation}
With this in hand, one can conclude that there additionally exists $\bar{w}>0$ (which should be chosen before one fixes $\bar{a}$) such that
\begin{align}
	\left|1-\bra{\psi}\left[\sum_{|a|\leq\bar{a}}\ket{a}\bra{a}\sum_{|A_{w}(\phi|\psi)|\leq\bar{w}}\ket{\phi}\bra{\phi}\right]\ket{\psi}\right| &
	\begin{aligned}[t]
		& = \left|1 - \braket{\psi|\psi(\bar{w})} + \braket{\psi|\psi(\bar{w})} - \sum_{|a|\leq\bar{a}}\braket{\psi|a}\braket{a|\psi(\bar{w})}\right| \\
		& \leq \left|1 - \braket{\psi|\psi(\bar{w})}\right| + \left|\braket{\psi|\psi(\bar{w})} - \sum_{|a|\leq\bar{a}}\braket{\psi|a}\braket{a|\psi(\bar{w})}\right| \\
		& < \xi.
	\end{aligned}
\end{align}
By extension, this implies that the real and imaginary components of Eqn. \ref{eq:Inner} converge, so for any $\xi>0$, there exist $\bar{a}>0$ and $\bar{w}>0$ such that
\begin{equation}
	\left|\Re\bra{\psi}\left[\sum_{|a|\leq\bar{a}}\ket{a}\bra{a}\sum_{|A_{w}(\phi|\psi)|\leq\bar{w}}\ket{\phi}\bra{\phi}\right]\ket{\psi} - 1\right| < \xi, \label{eq:Resumsmall}
\end{equation}
and
\begin{equation}
	\left|\Im\bra{\psi}\left[\sum_{|a|\leq\bar{a}}\ket{a}\bra{a}\sum_{|A_{w}(\phi|\psi)|\leq\bar{w}}\ket{\phi}\bra{\phi}\right]\ket{\psi}\right| < \xi. \label{eq:Imsumsmall}
\end{equation}
Once $\bar{a}$ and $\bar{w}$ are chosen such that Eqns. \ref{eq:Asum0} and \ref{eq:Wsum0} are made negligibly small and Eqns. \ref{eq:Resumsmall} and \ref{eq:Imsumsmall} are satisfied, $\epsilon$ can be chosen such that Eqns. \ref{eq:XiExpCos} and \ref{eq:XiSin} are satisfied. Then, Eqn. \ref{eq:NeumannDiffProgressed} can become
\begin{align}
	\left|\left|\sum_{|a|\leq\bar{a}}\ket{a}\braket{a|\psi}\ket{\mathcal{P}_{a\epsilon}}-\sum_{|A_{w}(\phi|\psi)|\leq\bar{w}}\ket{\phi}\braket{\phi|\psi}e^{i(\alpha+i\beta)\hat{q}-\Delta^{2}\beta^{2}}\ket{\mathscr{Q}}\right|\right|^{2} < 2 - 2(1-\xi) + 2\xi(1+\xi) + 2\xi^{2} = 4\xi(1+\xi).
\end{align}

Thus, for any $\xi > 0$, with proper choice of $\bar{a},\ \bar{w}$, and $\epsilon$, Eqn. \ref{eq:PrimeDiff} for the von Neumann model becomes
\begin{equation}
	\left|\left|e^{-i\frac{\epsilon}{\hbar}\hat{A}\hat{q}}\ket{\psi}\ket{\mathscr{Q}}-\sum_{\phi}\ket{\phi}\braket{\phi|\psi}\ket{\mathscr{Q}^{w}}\right|\right| < \xi(6+4\xi),
\end{equation}
where $\xi(6+4\xi)$ can be rectified (either through more careful choices in prior steps, or through definition of a new constant at the end) to a single arbitrarily small constant. In direct terms, it is possible to conclude that, as $\epsilon$ becomes sufficiently small, $e^{i\frac{\epsilon}{\hbar}\hat{A}\hat{q}}\ket{\psi}\ket{\mathscr{Q}}$ converges to $\sum_{\phi}\ket{\phi}\braket{\phi|\psi}\ket{\mathscr{Q}^{w}}$. In the wibbly wobbly way a proper physicist would write it, we conclude that
\begin{equation}
	e^{i\frac{\epsilon}{\hbar}\hat{A}\hat{q}}\ket{\psi}\ket{\mathscr{Q}} \approx \sum_{\phi}\ket{\phi}\braket{\phi|\psi}\ket{\mathscr{Q}^{w}},
\end{equation}
and thus
\begin{equation}
	\ket{\phi}\bra{\phi}e^{i\frac{\epsilon}{\hbar}\hat{A}\hat{q}}\ket{\psi}\ket{\mathscr{Q}} \approx \ket{\phi}\braket{\phi|\psi}\ket{\mathscr{Q}^{w}},
	\label{eq:NeumannSolved}
\end{equation}
as is necessary for the result of a weak measurement to be reliable.

\subsection{A Qubit Probe \label{subsec:Qubit}}
A similar exercise can in principle be done for a two-state probe system, such as a spin-$\frac{1}{2}$ particle. The initial probe state $\ket{\pi}$ becomes $\ket{+}_{z}$, the probe observable becomes $\hat{\sigma}_{x}$ (as is the choice in \cite{Hofmann2021direct}), and therefore the normalization is
\begin{align}
	N &
	\begin{aligned}[t]
		& = \zbra{+}e^{-2\frac{\epsilon}{\hbar}\Im A_{w}(\phi|\psi)\hat{\sigma}_{x}}\ket{+}_{z}^{-\frac{1}{2}} \\
		& = \left[\frac{1}{\sqrt{2}}\zbra{+}e^{-2\frac{\epsilon}{\hbar}\Im A_{w}(\phi|\psi)\hat{\sigma}_{x}}\left(\ket{+}_{x}+\ket{-}_{x}\right)\right]^{-\frac{1}{2}} \\
		& = \left[\frac{1}{2}\left(e^{-2\frac{\epsilon}{\hbar}\Im A_{w}(\phi|\psi)}+e^{2\frac{\epsilon}{\hbar}\Im A_{w}(\phi|\psi)}\right)\right]^{-\frac{1}{2}} \\
		& = \left[\cosh\left(2\frac{\epsilon}{\hbar}\Im A_{w}(\phi|\psi)\right)\right]^{-\frac{1}{2}} \\
		& = \sqrt{\text{sech}\left(2\frac{\epsilon}{\hbar}\Im A_{w}(\phi|\psi)\right)}.
	\end{aligned}
	\label{eq:QubitNorm}
\end{align}
Thus, the normalized probe state after the weak interaction is
\begin{equation}
	\ket{+^{w}}_{z} = \sqrt{\text{sech}\left(2\frac{\epsilon}{\hbar}\Im A_{w}(\phi|\psi)\right)}e^{i\frac{\epsilon}{\hbar}A_{w}(\phi|\psi)\hat{\sigma}_{x}}\ket{+}_{z},
\end{equation}
and the probability distributions of the spin-up and spin-down outcomes are
\begin{align}
	\mathscr{P}(+z) &
	\begin{aligned}[t]
		& = \left|\zbraket{+|+^{w}}_{z}\right|^{2} \\
		& = \text{sech}\left(2\frac{\epsilon}{\hbar}\Im A_{w}(\phi|\psi)\right)\left|\zbra{+}e^{i\frac{\epsilon}{\hbar}A_{w}(\phi|\psi)\hat{\sigma}_{x}}\ket{+}_{z}\right|^{2} \\
		& = \text{sech}\left(2\frac{\epsilon}{\hbar}\Im A_{w}(\phi|\psi)\right)\frac{1}{4}\left|e^{i\frac{\epsilon}{\hbar}A_{w}(\phi|\psi)} + e^{-i\frac{\epsilon}{\hbar}A_{w}(\phi|\psi)}\right|^{2} \\
		& = \text{sech}\left(2\frac{\epsilon}{\hbar}\Im A_{w}(\phi|\psi)\right)\frac{1}{4}\left(e^{i\frac{\epsilon}{\hbar}A_{w}(\phi|\psi)} + e^{-i\frac{\epsilon}{\hbar}A_{w}(\phi|\psi)}\right)\left(e^{-i\frac{\epsilon}{\hbar}A_{w}^{*}(\phi|\psi)} + e^{i\frac{\epsilon}{\hbar}A_{w}^{*}(\phi|\psi)}\right) \\
		& = \text{sech}\left(2\frac{\epsilon}{\hbar}\Im A_{w}(\phi|\psi)\right)\frac{1}{2}\left[\cosh\left(2\frac{\epsilon}{\hbar}\Im A_{w}(\phi|\psi)\right)+\cos\left(2\frac{\epsilon}{\hbar}\Re A_{w}(\phi|\psi)\right)\right] \\
		& = \frac{1}{2}\left[1+\cos\left(2\frac{\epsilon}{\hbar}\Re A_{w}(\phi|\psi)\right)\text{sech}\left(2\frac{\epsilon}{\hbar}\Im A_{w}(\phi|\psi)\right)\right],
	\end{aligned} \label{eq:Pplusz}
\end{align}
and
\begin{align}
	\mathscr{P}(-z) &
	\begin{aligned}[t]
		& = \left|\zbraket{-|+^{w}}_{z}\right|^{2} \\
		& = \text{sech}\left(2\frac{\epsilon}{\hbar}\Im A_{w}(\phi|\psi)\right)\frac{1}{4}\left|e^{i\frac{\epsilon}{\hbar}A_{w}(\phi|\psi)} - e^{-i\frac{\epsilon}{\hbar}A_{w}(\phi|\psi)}\right|^{2} \\
		& = \text{sech}\left(2\frac{\epsilon}{\hbar}\Im A_{w}(\phi|\psi)\right)\frac{1}{2}\left[\cosh\left(2\frac{\epsilon}{\hbar}\Im A_{w}(\phi|\psi)\right)-\cos\left(2\frac{\epsilon}{\hbar}\Re A_{w}(\phi|\psi)\right)\right] \\
		& = \frac{1}{2}\left[1-\cos\left(2\frac{\epsilon}{\hbar}\Re A_{w}(\phi|\psi)\right)\text{sech}\left(2\frac{\epsilon}{\hbar}\Im A_{w}(\phi|\psi)\right)\right].
	\end{aligned} \label{eq:Pminusz}
\end{align}
It is interesting to note that, where the von Neumann probe separated the real and imaginary parts of the weak value---one showing up in its momentum space probability density, and one showing up in its position space probability density, respectively---this setup for the qubit probe does not. The probability distribution for the $z$-component of spin is determined by both the real and imaginary parts.

It may be worth an aside to consider whether it is possible to separate the real and imaginary parts of the weak value in some basis. A brief discussion of this awaits in Appendix \ref{ap:QubitSep}.

Moving on, Eqn. \ref{eq:piapiw} becomes
\begin{align}
	\zbraket{+_{a}|+^{w}}_{z} &
	\begin{aligned}[t]
		& = \sqrt{\text{sech}\left(2\frac{\epsilon}{\hbar}\Im A_{w}(\phi|\psi)\right)}\zbra{+}e^{i\frac{\epsilon}{\hbar}\left[A_{w}(\phi|\psi)-a\right]\hat{\sigma}_{x}}\ket{+}_{z} \\
		& = \sqrt{\text{sech}\left(2\frac{\epsilon}{\hbar}\Im A_{w}(\phi|\psi)\right)} \frac{1}{2}\left[e^{i\frac{\epsilon}{\hbar}\left[A_{w}(\phi|\psi)-a\right]}+e^{-i\frac{\epsilon}{\hbar}\left[A_{w}(\phi|\psi)-a\right]}\right] \\
		& = \frac{1}{2}\sqrt{\text{sech}\left(2\frac{\epsilon}{\hbar}\Im A_{w}(\phi|\psi)\right)} \left[e^{i\frac{\epsilon}{\hbar}\left[\Re A_{w}(\phi|\psi)-a\right]}e^{-\frac{\epsilon}{\hbar}\Im A_{w}(\phi|\psi)}+e^{-i\frac{\epsilon}{\hbar}\left[\Re A_{w}(\phi|\psi)-a\right]}e^{\frac{\epsilon}{\hbar}\Im A_{w}(\phi|\psi)}\right] \\
		& = \sqrt{\text{sech}\left(2\frac{\epsilon}{\hbar}\Im A_{w}(\phi|\psi)\right)}\left[\cos\left(\frac{\epsilon}{\hbar}\left[\Re A_{w}(\phi|\psi)-a\right]\right)\cosh\left(\frac{\epsilon}{\hbar}\Im A_{w}(\phi|\psi)\right)\right. \\
		& \phantom{sqrt{\text{sech}\left(2\frac{\epsilon}{\hbar}\Im A_{w}(\phi|\psi)\right)}}-\left. i\sin\left(\frac{\epsilon}{\hbar}\left[\Re A_{w}(\phi|\psi)-a\right]\right)\sinh\left(\frac{\epsilon}{\hbar}\Im A_{w}(\phi|\psi)\right)\right].
	\end{aligned}
	\label{eq:plusaplusw}
\end{align}
Once again, we carry forward Eqn. \ref{eq:GeneralDiff}, which in the current notation, is
\begin{align}
	\left|\left|\sum_{|a|\leq\bar{a}}\ket{a}\braket{a|\psi}\ket{+_{a}}_{z} - \sum_{|A_{w}(\phi|\psi)|\leq \bar{w}}\ket{\phi}\braket{\phi|\psi}\ket{+^{w}}_{z} \right|\right|^{2} &
	\begin{aligned}[t]
		& = \sum_{|a|\leq\bar{a}}|\braket{a|\psi}|^{2} + \sum_{|A_{w}(\phi|\psi)|\leq \bar{w}}|\braket{\phi|\psi}|^{2} \\
		& - \sum_{|a|\leq\bar{a}} \sum_{|A_{w}(\phi|\psi)|\leq\bar{w}} \braket{\psi|a}\braket{a|\phi}\braket{\phi|\psi}\zbraket{+_{a}|+^{w}}_{z} \\
		& - \sum_{|a|\leq\bar{a}} \sum_{|A_{w}(\phi|\psi)|\leq\bar{w}} \braket{\psi|\phi}\braket{\phi|a}\braket{a|\psi}\zbraket{+^{w}|+_{a}}_{z}.
	\end{aligned}
	\label{eq:QubitDiff}
\end{align}
The last two terms can be rewritten as
\begin{align}
	\begin{aligned}[t]
		& -\sum_{|a|\leq\bar{a}} \sum_{|A_{w}(\phi|\psi)|\leq\bar{w}} \left(\braket{\psi|a}\braket{a|\phi}\braket{\phi|\psi}\zbraket{+_{a}|+^{w}}_{z} + \braket{\psi|\phi}\braket{\phi|a}\braket{a|\psi}\zbraket{+^{w}|+_{a}}_{z}\right) \\
		& = -2\sum_{|a|\leq\bar{a}} \sum_{|A_{w}(\phi|\psi)|\leq\bar{w}} \Re\left\{\braket{\psi|a}\braket{a|\phi}\braket{\phi|\psi}\zbraket{+_{a}|+^{w}}_{z}\right\},
	\end{aligned}
	\label{eq:QDoubSum}
\end{align}
and the summand can be expressed as
\begin{align}
	\begin{aligned}[t]
		& \Re\left\{\braket{\psi|a}\braket{a|\phi}\braket{\phi|\psi}\zbraket{+_{a}|+^{w}}_{z}\right\} \\
		& = \sqrt{\text{sech}\left(2\frac{\epsilon}{\hbar}\Im A_{w}(\phi|\psi)\right)}\left[\cos\left(\frac{\epsilon}{\hbar}\left[\Re A_{w}(\phi|\psi)-a\right]\right)\cosh\left(\frac{\epsilon}{\hbar}\Im A_{w}(\phi|\psi)\right)\Re\braket{\psi|a}\braket{a|\phi}\braket{\phi|\psi}\right. \\
		& \phantom{= \sqrt{\text{sech}\left(2\frac{\epsilon}{\hbar}\Im A_{w}(\phi|\psi)\right)}} \left.+\sin\left(\frac{\epsilon}{\hbar}\left[\Re A_{w}(\phi|\psi)-a\right]\right)\sinh\left(\frac{\epsilon}{\hbar}\Im A_{w}(\phi|\psi)\right)\Im\braket{\psi|a}\braket{a|\phi}\braket{\phi|\psi}\right],
	\end{aligned}
	\label{eq:QRPaQw}
\end{align}
The bounds on the real and imaginary parts of the weak value and the eigenvalues from Eqn. \ref{eq:BetweenPies} still hold, so it is possible to bound the functions in this expression:
\begin{align}
	\sqrt{\text{sech}\left(2\frac{\epsilon}{\hbar}\bar{w}\right)}\cos\left(\frac{\epsilon}{\hbar}\left[\bar{w}+\bar{a}\right]\right) & \leq \sqrt{\text{sech}\left(2\frac{\epsilon}{\hbar}\Im A_{w}(\phi|\psi)\right)}\cos\left(\frac{\epsilon}{\hbar}\left[\Re A_{w}(\phi|\psi)-a\right]\right)\cosh\left(\frac{\epsilon}{\hbar}\Im A_{w}(\phi|\psi)\right) \leq \cosh\left(\frac{\epsilon}{\hbar}\bar{w}\right), \label{eq:CoshSandwich}\\
	-\sinh\left(\frac{\epsilon}{\hbar}\bar{w}\right) & \leq \sqrt{\text{sech}\left(2\frac{\epsilon}{\hbar}\Im A_{w}(\phi|\psi)\right)}\sin\left(\frac{\epsilon}{\hbar}\left[\Re A_{w}(\phi|\psi)-a\right]\right)\sinh\left(\frac{\epsilon}{\hbar}\Im A_{w}(\phi|\psi)\right) \leq \sinh\left(\frac{\epsilon}{\hbar}\bar{w}\right). \label{eq:SinhSandwich}
\end{align}
For fixed $\bar{w}$ and $\bar{a}$, decreasing $\epsilon$ causes the lower and upper bounds of Eqn. \ref{eq:CoshSandwich} to approach 1, while the bounds of Eqn. \ref{eq:SinhSandwich} approach zero, squeezing the middle expressions between their respective bounds in each case. As such, for all $\xi>0$, and for all $\bar{w}>0$ and $\bar{a}>0$, there exists $\epsilon>0$ such that
\begin{align}
	\left|1-\sqrt{\text{sech}\left(2\frac{\epsilon}{\hbar}\Im A_{w}(\phi|\psi)\right)}\cos\left(\frac{\epsilon}{\hbar}\left[\Re A_{w}(\phi|\psi)-a\right]\right)\cosh\left(\frac{\epsilon}{\hbar}\Im A_{w}(\phi|\psi)\right)\right| & < \xi, \\
	\left|\sqrt{\text{sech}\left(2\frac{\epsilon}{\hbar}\Im A_{w}(\phi|\psi)\right)}\sin\left(\frac{\epsilon}{\hbar}\left[\Re A_{w}(\phi|\psi)-a\right]\right)\sinh\left(\frac{\epsilon}{\hbar}\Im A_{w}(\phi|\psi)\right)\right| & < \xi.
\end{align}
From here, the proof that Eqn. \ref{eq:QubitDiff} can be made negligibly small proceeds as in Section \ref{subsec:Neumann}, from Eqn. \ref{eq:NeumannDiffProgressed} onward, allowing one to conclude that $e^{i\frac{\epsilon}{\hbar}\hat{A}\hat{\sigma}_{x}}\ket{\psi}\ket{+}_{z}$ converges to $\sum_{\phi}\ket{\phi}\braket{\phi|\psi}\ket{+^{w}}_{z}$ as $\epsilon$ becomes sufficiently small. In other words,
\begin{equation}
	e^{i\frac{\epsilon}{\hbar}\hat{A}\hat{\sigma}_{x}}\ket{\psi}\ket{+}_{z} \approx \sum_{\phi}\ket{\phi}\braket{\phi|\psi}\ket{+^{w}}_{z},
\end{equation}
and thus
\begin{equation}
	\ket{\phi}\bra{\phi}e^{i\frac{\epsilon}{\hbar}\hat{A}\hat{\sigma}_{x}}\ket{\psi}\ket{+}_{z} \approx \ket{\phi}\braket{\phi|\psi}\ket{+^{w}}_{z}.
	\label{eq:QubitSolved}
\end{equation}

\section{Epilogue \label{sec:Epilogue}}
\subsection{Epilogue Proper}
In closing, I have demonstrated that, for $\epsilon$ sufficiently small, the exact state of a system undergoing a weak interaction with a probe,
\begin{equation}
	e^{i\frac{\epsilon}{\hbar}\hat{A}\hat{D}}\ket{\psi}\ket{\pi},
\end{equation}
converges to a superposition of entangled postselection states and (normalized) probe states, 
\begin{equation}
	\sum_{\phi}\ket{\phi}\braket{\phi|\psi}Ne^{i\frac{\epsilon}{\hbar}A_{w}(\phi|\psi)\hat{D}}\ket{\pi},
\end{equation}
thus validating the weak value approximation---a feat itself approximated, yet not truly performed, by prior calculations in the literature. I had to restrict myself to two specific measurement models---the von Neumann model ($\hat{D}=\hat{q}$ and $\ket{\pi}=\ket{\mathscr{Q}}$ as in Section \ref{subsec:Neumann}) and the qubit model ($\hat{D}=\hat{\sigma}_{x}$ and $\ket{\pi}=\ket{+}_{z}$ as in Section \ref{subsec:Qubit})---but otherwise, the assumptions (namely nondegeneracy of the observables $\hat{A}$ and $\hat{D}$, and that $|\braket{\phi|A|\psi}|<\infty$ for all $\ket{\phi}$) were minimal.

Additionally, by demonstrating a technique for proving convergence in these two cases, I have laid out a general foundation on which to build such derivations for other models in the same vein. This clarifies the path to validating the weak value approximation in schemes not yet considered, and there remains the possibility to redouble efforts on paths I have already tread so that the lingering assumptions may be weakened.

In the due course of strange aeons, the weak value approximation has found solid footing beyond the mire. Mayhap our slumber shall be less fitful, as forth from the single-paragraph, ersatz justifications of the elder days has hatched a more than two-dozen page monstrosity, here to allay our qualms.

\subsection{Where Credit is Due}
I would like to extend my gratitude to all those who read and gave perspectives on drafts of this paper. I received valuable input from members of my family, and from my advisor, Dr. David Craig. Let the record show that he warned strongly against the informal stylistic choices I made in my writing, but respected my choice to experiment and face the consequences.

As this is my first scientific paper, I would like to take the opportunity to thank all the teachers that I have ever had, including my first physics teacher, Mr. Lindstrom. Perhaps if I had listened to Mr. Larsen and majored in music, I would be less stressed out right now, but so it goes.

I would also like to thank my first reviewers for their efforts in helping me improve this paper. I especially appreciate Reviewer 1 for their encouraging comments about my writing style. Reviewer 3 and the editor did not agree with your assessment, however, so this version, with most of my informality intact, is for you.

\appendix
\section{Separating Imagination from Reality \label{ap:QubitSep}}
To attempt the separation of real and imaginary parts in the qubit probe, I begin by taking a general qubit state, the spin-up state along the vector $\hat{n}$ with azimuthal angle $\eta$ and polar angle $\theta$ \cite{McIntyre2012Quantum}:
\begin{equation}
	\ket{+}_{\hat{n}} = \cos\left(\frac{\theta}{2}\right)\ket{+}_{z} + e^{i\eta}\sin\left(\frac{\theta}{2}\right)\ket{-}_{z}.
\end{equation}
The amplitude of the $\ket{+}_{\hat{n}}$ component of $\ket{+^{w}}_{z}$ is
\begin{align}
	\begin{aligned}[t]
		& \prescript{}{\hat{n}}{\bra{+}}\sqrt{\text{sech}\left(2\frac{\epsilon}{\hbar}\Im A_{w}(\phi|\psi)\right)}e^{i\frac{\epsilon}{\hbar}A_{w}(\phi|\psi)\hat{\sigma}_{x}}\ket{+}_{z} \\
		& = \frac{1}{\sqrt{2}}\sqrt{\text{sech}\left(2\frac{\epsilon}{\hbar}\Im A_{w}(\phi|\psi)\right)}\prescript{}{\hat{n}}{\bra{+}}\left(e^{i\frac{\epsilon}{\hbar}A_{w}(\phi|\psi)}\ket{+}_{x}+e^{-i\frac{\epsilon}{\hbar}A_{w}(\phi|\psi)}\ket{-}_{x}\right) \\
		& = \frac{1}{2}\sqrt{\text{sech}\left(2\frac{\epsilon}{\hbar}\Im A_{w}(\phi|\psi)\right)}\prescript{}{\hat{n}}{\bra{+}} \left[\left(e^{i\frac{\epsilon}{\hbar}A_{w}(\phi|\psi)}+e^{-i\frac{\epsilon}{\hbar}A_{w}(\phi|\psi)}\right)\ket{+}_{z}+\left(e^{i\frac{\epsilon}{\hbar}A_{w}(\phi|\psi)}-e^{-i\frac{\epsilon}{\hbar}A_{w}(\phi|\psi)}\right)\ket{-}_{z}\right] \\
		& = \frac{1}{2}\sqrt{\text{sech}\left(2\frac{\epsilon}{\hbar}\Im A_{w}(\phi|\psi)\right)} \left[\left(e^{i\frac{\epsilon}{\hbar}A_{w}(\phi|\psi)}+e^{-i\frac{\epsilon}{\hbar}A_{w}(\phi|\psi)}\right)\cos\left(\frac{\theta}{2}\right)+\left(e^{i\frac{\epsilon}{\hbar}A_{w}(\phi|\psi)}-e^{-i\frac{\epsilon}{\hbar}A_{w}(\phi|\psi)}\right)e^{-i\eta}\sin\left(\frac{\theta}{2}\right)\right].
	\end{aligned}
\end{align}
This leads us to the probability of finding the qubit to be in the spin-up state along the $\hat{n}$ axis, which is
\begin{align}
	\mathscr{P}(+\hat{n})
	\begin{aligned}[t]
		& = \left|\prescript{}{\hat{n}}{\bra{+}}\sqrt{\text{sech}\left(2\frac{\epsilon}{\hbar}\Im A_{w}(\phi|\psi)\right)}e^{i\frac{\epsilon}{\hbar}A_{w}(\phi|\psi)\hat{\sigma}_{x}}\ket{+}_{z}\right|^{2} \\
		& = \frac{1}{4}\text{sech}\left(2\frac{\epsilon}{\hbar}\Im A_{w}(\phi|\psi)\right) \left[\cos^{2}\left(\frac{\theta}{2}\right)\left(e^{i\frac{\epsilon}{\hbar}A_{w}(\phi|\psi)}+e^{-i\frac{\epsilon}{\hbar}A_{w}(\phi|\psi)}\right)\left(e^{-i\frac{\epsilon}{\hbar}A_{w}^{*}(\phi|\psi)}+e^{i\frac{\epsilon}{\hbar}A_{w}^{*}(\phi|\psi)}\right)\right. \\
		& + \cos\left(\frac{\theta}{2}\right)\sin\left(\frac{\theta}{2}\right)\left(e^{i\frac{\epsilon}{\hbar}A_{w}(\phi|\psi)}+e^{-i\frac{\epsilon}{\hbar}A_{w}(\phi|\psi)}\right)\left(e^{-i\frac{\epsilon}{\hbar}A_{w}^{*}(\phi|\psi)}-e^{i\frac{\epsilon}{\hbar}A_{w}^{*}(\phi|\psi)}\right)e^{i\eta} \\
		& + \cos\left(\frac{\theta}{2}\right)\sin\left(\frac{\theta}{2}\right)\left(e^{i\frac{\epsilon}{\hbar}A_{w}(\phi|\psi)}-e^{-i\frac{\epsilon}{\hbar}A_{w}(\phi|\psi)}\right)\left(e^{-i\frac{\epsilon}{\hbar}A_{w}^{*}(\phi|\psi)}+e^{i\frac{\epsilon}{\hbar}A_{w}^{*}(\phi|\psi)}\right)e^{-i\eta} \\
		& + \left.\sin^{2}\left(\frac{\theta}{2}\right)\left(e^{i\frac{\epsilon}{\hbar}A_{w}(\phi|\psi)}-e^{-i\frac{\epsilon}{\hbar}A_{w}(\phi|\psi)}\right)\left(e^{-i\frac{\epsilon}{\hbar}A_{w}^{*}(\phi|\psi)}-e^{i\frac{\epsilon}{\hbar}A_{w}^{*}(\phi|\psi)}\right)\right] \\
		& = \frac{1}{4}\text{sech}\left(2\frac{\epsilon}{\hbar}\Im A_{w}(\phi|\psi)\right) \left[\cos^{2}\left(\frac{\theta}{2}\right)\left(2\cosh\left(2\frac{\epsilon}{\hbar}\Im A_{w}(\phi|\psi)\right)+2\cos\left(2\frac{\epsilon}{\hbar}\Re A_{w}(\phi|\psi)\right)\right)\right. \\
		& - \cos\left(\frac{\theta}{2}\right)\sin\left(\frac{\theta}{2}\right)e^{i\eta}\left(2\sinh\left(2\frac{\epsilon}{\hbar}\Im A_{w}(\phi|\psi)\right)+2i\sin\left(2\frac{\epsilon}{\hbar}\Re A_{w}(\phi|\psi)\right)\right) \\
		& - \cos\left(\frac{\theta}{2}\right)\sin\left(\frac{\theta}{2}\right)e^{-i\eta}\left(2\sinh\left(2\frac{\epsilon}{\hbar}\Im A_{w}(\phi|\psi)\right)-2i\sin\left(2\frac{\epsilon}{\hbar}\Re A_{w}(\phi|\psi)\right)\right) \\
		& + \left.\sin^{2}\left(\frac{\theta}{2}\right)\left(2\cosh\left(2\frac{\epsilon}{\hbar}\Im A_{w}(\phi|\psi)\right)-2\cos\left(2\frac{\epsilon}{\hbar}\Re A_{w}(\phi|\psi)\right)\right)\right] \\
		& = \frac{1}{4}\text{sech}\left(2\frac{\epsilon}{\hbar}\Im A_{w}(\phi|\psi)\right) \left[2\cosh\left(2\frac{\epsilon}{\hbar}\Im A_{w}(\phi|\psi)\right)+2\cos(\theta)\cos\left(2\frac{\epsilon}{\hbar}\Re A_{w}(\phi|\psi)\right)\right. \\
		& - 2\sin\left(\theta\right)\cos(\eta)\sinh\left(2\frac{\epsilon}{\hbar}\Im A_{w}(\phi|\psi)\right) + 2\sin\left(\theta\right)\sin(\eta)\sin\left(2\frac{\epsilon}{\hbar}\Re A_{w}(\phi|\psi)\right) \\
		& = \frac{1}{2}\left[1+\cos(\theta)\cos\left(2\frac{\epsilon}{\hbar}\Re A_{w}(\phi|\psi)\right)\text{sech}\left(2\frac{\epsilon}{\hbar}\Im A_{w}(\phi|\psi)\right)\right. \\
		&+\left.\sin(\theta)\left(\sin\left(2\frac{\epsilon}{\hbar}\Re A_{w}(\phi|\psi)\right)\text{sech}\left(2\frac{\epsilon}{\hbar}\Im A_{w}(\phi|\psi)\right)\sin(\eta)-\tanh\left(2\frac{\epsilon}{\hbar}\Im A_{w}(\phi|\psi)\right)\cos(\eta)\right)\right].
	\end{aligned}
\end{align}
It is immediately apparent that setting $\theta=\frac{\pi}{2}$ and $\eta=0$ (that is, setting $\ket{+}_{\hat{n}}=\ket{+}_{x}$) gives one
\begin{equation}
	\mathscr{P}(+\hat{n})=\frac{1}{2}\left[1-\tanh\left(2\frac{\epsilon}{\hbar}\Im A_{w}(\phi|\psi)\right)\right],
\end{equation}
allowing the experimenter to find only the imaginary part of the weak value by examining the statistics of the probe outcomes. This is enough, as one can learn the imaginary part to high accuracy with this measurement, then account for it in the $z$-basis probabilities calculated in Eqns. \ref{eq:Pplusz} and \ref{eq:Pminusz}, and subsequently determine the real part in another round of $z$-basis measurements. It is a good thing that this is possible, as there does not appear to be a similar way to isolate the real component. Consider (taking our convention from Eqn. \ref{eq:AlphaBeta}) that
\begin{equation}
	\parn{}{\beta}{}\mathscr{P}(+\hat{n}) = -\text{sech}^{2}(2\beta)\left[\cos(\theta)\cos(2\alpha)\sinh(2\beta)+\sin(\theta)\left(\sin(2\alpha)\sin(\eta)\sinh(2\beta)+\cos(\eta)\right)\right].
\end{equation}
Our goal would be to find $\theta$ and $\eta$ such that $\mathscr{P}(+\hat{n})$ is independent of the imaginary part of the weak value (which would mean its derivative with respect to $\beta$ is zero), so in general, we want to show that
\begin{equation}
	0 = \left[\cos(\theta)\cos(2\alpha)+\sin(\theta)\sin(2\alpha)\sin(\eta)\right]\sinh(2\beta)+\sin(\theta)\cos(\eta),
\end{equation}
which is true if and only if
\begin{equation}
	\cos(\theta)\cos(2\alpha)+\sin(\theta)\sin(2\alpha)\sin(\eta)=-\sin(\theta)\cos(\eta)\text{csch}(2\beta).
\end{equation}
By separation of variables, one can see that, if this is to be true for all $\alpha$ and $\beta$, then each side must be equal to zero. Furthermore, the left cannot be zero for all $\alpha$ if $\sin(\theta)=0$, so it must be that $\cos(\eta)=0$. This means $\sin(\eta)=\pm1$, which leaves the requirement
\begin{equation}
	0=\cos(\theta)\cos(2\alpha)\pm\sin(\theta)\sin(2\alpha).
\end{equation}
If one were to separate variables again, one would see that this is not possible for a general $\alpha$ if $\theta$ does not depend on it.

\section{Alternative Approach to the Weak Value Approximation \label{ap:Agnostic}}
For my next trick, we return to Eqn. \ref{eq:AjwAwjDiff}. Now, that was written in a manner consistent with prior literature, so I will first make a revision that incorporates the properly normalized probe state, as defined in Eqn. \ref{eq:Normalize}:
\begin{align}
	\left|\left|\ket{\phi}\bra{\phi}e^{i\frac{\epsilon}{\hbar}\hat{A}\hat{D}}\ket{\psi}\ket{\pi}- \ket{\phi}\braket{\phi|\psi}Ne^{i\frac{\epsilon}{\hbar}A_{w}(\phi|\psi)\hat{D}}\ket{\pi}\right|\right| &
	\begin{aligned}[t]
		& = \left|\left|\ket{\phi}\braket{\phi|\psi}\sum_{j=0}^{\infty}\frac{1}{j!}\left(i\frac{\epsilon}{\hbar}\hat{D}\right)^{j}\left[\left(\hat{A}^{j}\right)_{w}-N\left(A_{w}\right)^{j}\right]\ket{\pi}\right|\right| \\
		& = |\braket{\phi|\psi}|\left|\left|\sum_{j=0}^{\infty}\frac{1}{j!}\left(i\frac{\epsilon}{\hbar}\hat{D}\right)^{j}\left[\left(\hat{A}^{j}\right)_{w}-N\left(A_{w}\right)^{j}\right]\ket{\pi}\right|\right| \\
		& \leq |\braket{\phi|\psi}|\sum_{j=0}^{\infty}\frac{1}{j!}\left(\frac{\epsilon}{\hbar}\right)^{j}\left|\left(\hat{A}^{j}\right)_{w}-N\left(A_{w}\right)^{j}\right|\left|\left|\hat{D}^{j}\ket{\pi}\right|\right|
	\end{aligned}
	\label{eq:AjwAwjDiffNorm}
\end{align}
While the main focus will be on $\left|\left(\hat{A}^{j}\right)_{w}-N\left(A_{w}\right)^{j}\right|$, it is still necessary to know what sort of probe is being used in order to handle $\left|\left|\hat{D}^{j}\ket{\pi}\right|\right|$. Since \cite{PhysRevA.41.11} used the von Neumann model, that is what I will use. As such, it is necessary to calculate (for $j\in\mathbb{N}$)
\begin{equation}
	\begin{split}
		\left|\left|\hat{q}^{j}\ket{\mathscr{Q}}\right|\right| & = \sqrt{\braket{\mathscr{Q}|\hat{q}^{2j}|\mathscr{Q}}} \\
		& = \sqrt{\int_{-\infty}^{\infty}dq\int_{-\infty}^{\infty}dq'\frac{q'^{2j}}{\sqrt{2\pi\Delta^{2}}}\exp\left(\frac{-q^{2}-q'^{2}}{4\Delta^{2}}\right)\braket{q|q'}} \\
		& = \sqrt{\int_{-\infty}^{\infty}dq\frac{q^{2j}}{\sqrt{2\pi\Delta^{2}}}\exp\left(\frac{-q^{2}}{2\Delta^{2}}\right)} \\
		& = \sqrt{2\int_{0}^{\infty}dq\frac{q^{2j}}{\sqrt{2\pi\Delta^{2}}}\exp\left(\frac{-q^{2}}{2\Delta^{2}}\right)}.
	\end{split}
\end{equation}
Making the substitution $ x = \frac{q^{2}}{2\Delta^{2}} $ allows one to obtain
\begin{equation}
	\begin{split}
		\left|\left|\hat{q}^{j}\ket{\mathscr{Q}}\right|\right| & = \sqrt{2\int_{0}^{\infty}dx\frac{\Delta^{2}(2\Delta^{2}x)^{\frac{2j-1}{2}}}{\sqrt{2\pi\Delta^{2}}}e^{-x}} \\
		& = \sqrt{\frac{(\sqrt{2}\Delta)^{2j}}{\sqrt{\pi}}\int_{0}^{\infty}dxx^{\left(j+\frac{1}{2}\right)-1}e^{-x}}.
	\end{split}
\end{equation}
I desire this form, because the gamma function is defined by $ \Gamma(j) = \int_{0}^{\infty}dxx^{j-1}e^{-x} $ \cite{Boas2006mathematical}, so that integral is $ \Gamma\left(j+\frac{1}{2}\right) $. Since $ \Gamma\left(j+\frac{1}{2}\right) = \frac{(2j-1)!!}{2^{j}}\sqrt{\pi} $ \cite{MathworldGammaFxn} (see Appendix \ref{ap:GammaWorld}), one finds
\begin{equation}
	\left|\left|\hat{q}^{j}\ket{\mathscr{Q}}\right|\right| = \Delta^{j} \sqrt{(2j-1)!!},
\end{equation}
at least, for $j>0$, as was previously stipulated above. However, we have a $j=0$ term, which must be separated out. With that out of the way, it can be concluded that
\begin{equation}
	\left|\left|\ket{\phi}\bra{\phi}e^{i\frac{\epsilon}{\hbar}\hat{A}\hat{q}}\ket{\psi}\ket{\mathscr{Q}}- \ket{\phi}\braket{\phi|\psi}Ne^{i\frac{\epsilon}{\hbar}A_{w}(\phi|\psi)\hat{q}}\ket{\mathscr{Q}}\right|\right| \leq |\braket{\phi|\psi}|\ |1-N|+|\braket{\phi|\psi}|\sum_{j=1}^{\infty}\frac{\sqrt{(2j-1)!!}}{j!}\left(\frac{\epsilon\Delta}{\hbar}\right)^{j}\left|\left(\hat{A}^{j}\right)_{w}-N\left(A_{w}\right)^{j}\right|.
	\label{eq:SumPig}
\end{equation}
As such, to validate the weak value approximation, I need to make the right hand side of Eqn. \ref{eq:SumPig} negligibly small, which requires bounding the sum. I can determine $ \Delta $, as the hypothetical experimenter controls the preparation of the probe state, and I have control over the coupling strength $ \epsilon $, as the experimenter controls the experimental setup. As such, the only two factors of concern are $ \frac{\sqrt{(2j-1)!!}}{j!} $ and $ \left|\left(\hat{A}^{j}\right)_{w}-N\left(A_{w}\right)^{j}\right| $.

Regarding the former, it can be shown (in Appendix \ref{ap:SubExp}) that, for any $ a \geq 1 $ there exists a natural number $ k $ such that $ \frac{\sqrt{(2j-1)!!}}{j!} < a^{-j} $ for all $ j \geq k $. This is an exceptional decay, stronger than any exponential growth, so as long as $ \left|(\hat{A}^{j})_{w}-A_{w}^{j}\right| $ does not grow at a more than exponential rate, the right hand side of Eqn. \ref{eq:SumPig} will converge. For now, let it suffice that $ \frac{\sqrt{(2j-1)!!}}{j!} \leq 1 $ for all $ j \in \mathbb{N} $.

Ultimately, I need more than just convergence, which requires me to know how $ \left|\left(\hat{A}^{j}\right)_{w}-N\left(A_{w}\right)^{j}\right| $ behaves. There are levels of complexity to this, the simplest being if either $ \ket{\psi} $ or $ \ket{\phi} $ is an eigenvector of $ \hat{A} $. In this case, $ A_{w} $ is simply the associated eigenvalue $a$ (unless of course $ \ket{\psi} $ and $ \ket{\phi} $ are orthogonal, in which case the probe state is not shifted). The weak value $ (\hat{A}^{j})_{w} $ is this eigenvalue to the $ j $th power, so $ \left|\left(\hat{A}^{j}\right)_{w}-N\left(A_{w}\right)^{j}\right| = |a|^{j}(1-N) = 0 $ (as $N=1$ when the imaginary part of the weak value is zero), and no approximation is necessary. Note that absolute value bars on $1-N$ are also unnecessary, as Eqn. \ref{eq:NormBound} tells us that $N\leq1$ (as $\braket{\mathscr{Q}|\hat{q}|\mathscr{Q}}=0$).

Increasing the complexity, let us consider bounded operators. The operator norm is defined by $ ||\hat{A}||_{op} = \sup_{||\ket{\psi}|| \leq 1}||\hat{A}\ket{\psi}|| $ (\cite{Axler2020Measure} page 167), and thus $ ||\hat{A}\ket{\psi}|| \leq ||\hat{A}||_{op}||\ket{\psi}|| $. This adds the convenience that $ ||\hat{A}^{n}\ket{\psi}|| \leq ||A||_{op}||\hat{A}^{n-1}\ket{\psi}|| \leq \dots \leq ||\hat{A}||_{op}^{n}||\ket{\psi}|| $. As such, I can apply the \ac{CSB} inequality to the absolute value of the weak value of $ \hat{A}^{n} $ to find
\begin{equation}
	\left|(\hat{A}^{n})_{w}\right| = \left|\frac{\braket{\phi|\hat{A}^{n}|\psi}}{\braket{\phi|\psi}}\right| \leq \frac{||\ket{\phi}||\ ||\hat{A}^{n}\ket{\psi}||}{|\braket{\phi|\psi}|} \leq ||\hat{A}||_{op}^{n}\frac{||\ket{\phi}||\ ||\ket{\psi}||}{|\braket{\phi|\psi}|}.
\end{equation}
As a result,
\begin{equation}
	\begin{split}
		\left|\left(\hat{A}^{j}\right)_{w}-N\left(A_{w}\right)^{j}\right| & \leq \left|(\hat{A}^{j})_{w}\right| + \left|A_{w}\right|^{j} \\
		& \leq ||\hat{A}||_{op}^{j}\frac{||\ket{\phi}||\ ||\ket{\psi}||}{|\braket{\phi|\psi}|} + ||\hat{A}||_{op}^{j}\left(\frac{||\ket{\phi}||\ ||\ket{\psi}||}{|\braket{\phi|\psi}|}\right)^{j} \\
		& = \left(\frac{1}{|\braket{\phi|\psi}|} + \frac{1}{|\braket{\phi|\psi}|^{j}}\right)||\hat{A}||_{op}^{j},
	\end{split}
\end{equation}
and since $ |\braket{\phi|\psi}| \leq 1 $ I know that $ \frac{1}{|\braket{\phi|\psi}|} \leq \frac{1}{|\braket{\phi|\psi}|^{j}} $ for all $ j \in \mathbb{N} $. This allows me to implement the (very loose, but conveniently simple) bound
\begin{equation}
	\left|\left(\hat{A}^{j}\right)_{w}-N\left(A_{w}\right)^{j}\right| \leq 2\left(\frac{||\hat{A}||_{op}}{|\braket{\phi|\psi}|}\right)^{j}.
\end{equation}
With this, Eqn. \ref{eq:SumPig} becomes
\begin{equation}
	\left|\left|\ket{\phi}\bra{\phi}e^{i\frac{\epsilon}{\hbar}\hat{A}\hat{q}}\ket{\psi}\ket{\mathscr{Q}}- \ket{\phi}\braket{\phi|\psi}Ne^{i\frac{\epsilon}{\hbar}A_{w}(\phi|\psi)\hat{q}}\ket{\mathscr{Q}}\right|\right| \leq 1-N+2\sum_{j=1}^{\infty}\left(\frac{\epsilon\Delta}{\hbar}\frac{||\hat{A}||_{op}}{|\braket{\phi|\psi}|}\right)^{j}. \label{eq:SumPork}
\end{equation}
thus, I have a geometric series, which I can shrink as much as I need. Let $ r = \frac{\epsilon\Delta}{\hbar}\frac{||\hat{A}||_{op}}{|\braket{\phi|\psi}|} < 1 $, and note that
\begin{equation}
	\sum_{j=1}^{\infty}r^{j} = \frac{1}{1-r}-1 = \frac{r}{1-r}.
\end{equation}
To make $ 2\sum_{j=1}^{\infty}r^{j} \ll 1 $, I require $ \frac{r}{1-r} \ll \frac{1}{2} $, which is achieved when $ r \ll \frac{1}{3} $. This tells me that I must choose $ \epsilon $ and $ \Delta $ such that
\begin{equation}
	\epsilon\Delta \ll \frac{|\braket{\phi|\psi}|}{||\hat{A}||_{op}}\frac{\hbar}{3}.
	\label{eq:OpBound}
\end{equation}
I also need to deal with bounding $1-N$ in Eqn. \ref{eq:SumPork}, but I will postpone that until the end.

Next, I have to deal with unbounded operators, which is a more complicated notion. Here, I have to use more knowledge about the input and output states, $ \ket{\psi} $ and $ \ket{\phi} $, which in turn divides into two cases: states with compact support in the state space and states without. The former case is tractable, but the latter case is too general for this method, at least in my hands.

Let us assume without loss of generality that $ \ket{\psi} $ has compact support, particularly when represented in the eigenvalue spectrum of $ \hat{A} $, denoted $\text{sp}(\hat{A}) $. Thus, 
\begin{equation}
	\ket{\psi} = \int_{a\in\text{sp}(\hat{A})}da\ \psi(a)\ket{a},
\end{equation}
and let $ \text{supp}(\psi(a)) = \{a \in \text{sp}(\hat{A}):\psi(a)\neq0\} $ be the support of the wave function $ \psi(a) $. It follows that
\begin{equation}
	\hat{A}^{n}\ket{\psi} = \int_{a\in\text{sp}(\hat{A})}da\ \psi(a)a^{n}\ket{a}.
\end{equation}
Since $ \text{supp}(\psi(a)) $ is compact, one can define $ a_{max} = \sup\{|a|:a\in\text{supp}(\psi(a))\} $, which will be a finite quantity, and thus
\begin{equation}
	\begin{split}
		||\hat{A}^{n}\ket{\psi}||^{2} & = \left|\left| \int_{a\in\text{sp}(\hat{A})}da\ \psi(a)a^{n}\ket{a} \right|\right|^{2} \\
		& = \int_{a'\in\text{sp}(\hat{A})}da'\int_{a\in\text{sp}(\hat{A})}da\ \psi^{*}(a')\psi(a)a'^{n}a^{n}\braket{a'|a} \\
		& = \int_{a\in\text{sp}(\hat{A})}da\ |\psi(a)|^{2}a^{2n} \\
		& \leq a_{max}^{2n} \int_{a\in\text{sp}(\hat{A})}da\ |\psi(a)|^{2} \\
		& = a_{max}^{2n} ||\ket{\psi}||^{2}.
	\end{split}
	\label{eq:Angband}
\end{equation}
As such, $ ||\hat{A}^{n}\ket{\psi}|| \leq a_{max}^{n} ||\ket{\psi}|| $, and so
\begin{equation}
	\left|\left(\hat{A}^{j}\right)_{w}-N\left(A_{w}\right)^{j}\right| \leq \left(\frac{1}{|\braket{\phi|\psi}|} + \frac{1}{|\braket{\phi|\psi}|^{j}}\right)a_{max}^{j}.
\end{equation}
This allows me to bound our error effectively by choosing $ \epsilon $ and $ \Delta $ such that
\begin{equation}
	\epsilon\Delta \ll \frac{|\braket{\phi|\psi}|}{a_{max}}\frac{\hbar}{3}.
	\label{eq:CompBound}
\end{equation}

Finally, as I promised, I must bound $1-N$. From Eqn. \ref{eq:NeumannNorm}, we know that
\begin{equation}
	1-N = 1-e^{-\frac{\epsilon^{2}\Delta^{2}}{\hbar^{2}}\left(\Im A_{w}\right)^{2}}
\end{equation}
If the weak value is real, then this difference is already zero, and my work is done. Thus, I shall assume the weak value is complex. For decreasing $\epsilon\Delta$, the exponential is monotone increasing toward a maximum of 1, so it must be possible to find $\epsilon\Delta$ such that this difference is negligibly small. If $\nu\ll1$ is the largest we want this difference to be (that is, we desire $\nu \geq 1-N$), then we require that
\begin{equation}
	\epsilon\Delta \leq \frac{\hbar}{|\Im A_{w}|}\sqrt{\ln\left(\frac{1}{1-\nu}\right)},
\end{equation}
which is finite so long as the imaginary part of the weak value is finite (which must be true for a bounded operator and an unbounded operator acting on a state with compact support). Taking the stricter of this condition and whichever condition above applies (Eqn. \ref{eq:OpBound} or Eqn. \ref{eq:CompBound}) validates the weak value approximation.

\section{Proof of $ \Gamma\left(j+\frac{1}{2}\right) = \frac{(2j-1)!!}{2^{j}}\sqrt{\pi} $\label{ap:GammaWorld}}
While the claim $ \Gamma\left(j+\frac{1}{2}\right) = \frac{(2j-1)!!}{2^{j}}\sqrt{\pi} $ for $ j \in \mathbb{N} $ is provided by \cite{MathworldGammaFxn}, the proof is omitted, likely due to its elementary nature. Those familiar with the gamma function know that $ \Gamma(x+1) = x\Gamma(x) $ (which is shown by integration by parts), so once the base case, $ \Gamma(1+\frac{1}{2}) = \frac{\sqrt{\pi}}{2} $, is proved, $ \Gamma\left(j+\frac{1}{2}\right) = \frac{(2j-1)!!}{2^{j}}\sqrt{\pi} $ follows immediately by induction. In particular,
\begin{equation}
	\Gamma\left(j+\frac{3}{2}\right) = \left(j+\frac{1}{2}\right)\Gamma\left(j+\frac{1}{2}\right) = \frac{2j+1}{2}\frac{(2j-1)!!}{2^{j}}\sqrt{\pi} = \frac{(2j+1)!!}{2^{j+1}}\sqrt{\pi}.
\end{equation}

The base case is also simple, requiring a little substitution and some knowledge of Gaussian integrals. I start with
\begin{equation}
	\Gamma\left(\frac{3}{2}\right) = \int_{0}^{\infty}dx\ \sqrt{x}e^{-x},
\end{equation}
and make the substitution $ z^{2} = x $, which gives me
\begin{equation}
	\begin{split}
		\Gamma\left(\frac{3}{2}\right) & = 2\int_{0}^{\infty}dz\ z^{2} e^{-z^{2}} \\
		& = \left[-\parn{}{t}{}\int_{-\infty}^{\infty}dz\ e^{-tz^{2}}\right]_{t=1} \\
		& = \left[-\parn{}{t}{}\sqrt{\frac{\pi}{t}}\ \right]_{t=1} \\
		& = \left[\frac{\sqrt{\pi}}{2}t^{-3/2}\right]_{t=1} \\
		& = \frac{\sqrt{\pi}}{2}.
	\end{split}
\end{equation}

\section{Limiting Behavior of $\frac{\sqrt{(2j-1)!!}}{j!}$\label{ap:SubExp}}
To prove that, for any $ a \geq 1 $ there exists a natural number $ k $ such that $ \frac{\sqrt{(2j-1)!!}}{j!} < a^{-j} $ for all $ j \geq k $, I start with a much simpler fact: for any $ a $, there exists $ k \in \mathbb{N} $ such that $ k \geq 2e^{16}a^{2} $. Now, for all $ j \geq k $, it follows that $ j \geq 2e^{16}a^{2} $, and I can multiply both sides by $ j^{2} $ (and play a bit with the right hand side) to get
\begin{equation}
	j^{3} \geq e^{16}\frac{a^{2}}{2}(2j)^{2}.
\end{equation}
If I take the logarithm of both sides, then
\begin{equation}
	3\ln j \geq 16 + \ln\frac{a^{2}}{2} + 2\ln2j.
\end{equation}
I can then multiply by $ j $ to get
\begin{equation}
	3j\ln j \geq 16j + j\ln\frac{a^{2}}{2} + 2j\ln2j.
\end{equation}
Now I can subtract $ 42j $ from each side to get
\begin{equation}
	3j\ln j - 42j \geq -26j + j\ln\frac{a^{2}}{2} + 2j\ln2j.
\end{equation}
Next, I have two more inequalities to work with. The first is a comparison of constants: $ 3\ln\sqrt{2\pi} > \ln\sqrt{2\pi} + \frac{1}{2}\ln2 $. Second, I know that $ \frac{3}{2}\ln j \geq \frac{1}{2}\ln j $ for all $ j \geq 1 $. As such, I can add these to our expression to get
\begin{equation}
	3j\ln j - 42j + 3\ln\sqrt{2\pi} + \frac{3}{2}\ln j \geq -26j + j\ln\frac{a^{2}}{2} + 2j\ln2j + \ln\sqrt{2\pi} + \frac{1}{2}\ln 2j.
	\label{eq:PenultimatePeril}
\end{equation}
Why is this important? Well, the factorial is bounded in the following manner \cite{Robbins1955Remark}:
\begin{equation}
	\sqrt{2\pi n}\left(\frac{n}{e}\right)^{n}\frac{1}{e^{12n+1}} < n! < \sqrt{2\pi n}\left(\frac{n}{e}\right)^{n}\frac{1}{e^{12n}}.
\end{equation}
As such, it follows that
\begin{align}
	\ln(j!)^{3} & > 3\ln\sqrt{2\pi} + 3j\ln j + \frac{3}{2}\ln j - 39j - 3; \label{eq:RemarkExt1}\\
	\ln\left(\frac{a^{2}}{2}\right)^{j}(2j)! & < j\ln\frac{a^{2}}{2} + \ln\sqrt{2\pi} + 2j\ln2j + \frac{1}{2}\ln2j - 26j. \label{eq:RemarkExt2}
\end{align}
As such,
\begin{equation}
	\begin{split}
		\ln(j!)^{3} & > 3\ln\sqrt{2\pi} + 3j\ln j + \frac{3}{2}\ln j - 39j - 3 \\
		& \geq 3\ln\sqrt{2\pi} + 3j\ln j + \frac{3}{2}\ln j - 39j - 3j \\
		& = 3\ln\sqrt{2\pi} + 3j\ln j + \frac{3}{2}\ln j - 42j  \\
		& \geq j\ln\frac{a^{2}}{2} + \ln\sqrt{2\pi} + 2j\ln2j + \frac{1}{2}\ln 2j - 26j \\
		& > \ln\left(\frac{a^{2}}{2}\right)^{j}(2j)!.
	\end{split}
\end{equation}
(To clarify the above, I started with Eqn. \ref{eq:RemarkExt1}, used Eqn. \ref{eq:PenultimatePeril} to go from line 3 to line 4, and finished using Eqn. \ref{eq:RemarkExt2}.)

Exponentiating both sides gives
\begin{equation}
	(j!)^{3} > \left(\frac{a^{2}}{2}\right)^{j}(2j)!,
\end{equation}
and since $ j! = \frac{(2j)!!}{2^{j}} $, the expression can be divided by this to obtain
\begin{equation}
	(j!)^{2} > a^{2j}(2j-1)!!.
\end{equation}
Now, I can take the square root and rearrange to find that
\begin{equation}
	\frac{\sqrt{(2j-1)!!}}{j!} < a^{-j}.
	\label{eq:ExceptionalBound}
\end{equation}
Thus, for any $ a \geq 1 $, we can select $ k \in \mathbb{N} $ such that Eqn. \ref{eq:ExceptionalBound} holds for all $ j \geq k $, with the easiest way to find this $ k $ being to choose it such that $ k \geq 2e^{16}a^{2} $ (a number well over 17 million).

Of course, less extreme upper bounds on $ \frac{\sqrt{(2j-1)!!}}{j!} $ require less extreme lower bounds on $ k $. In particular, assume that there exists $ j \in \mathbb{N} $ such that $ j! \geq \sqrt{(2j-1)!!} $. It follows that
\begin{equation}
	(j+1)! = (j+1)j! \geq (j+1)\sqrt{(2j-1)!!} = \frac{j+1}{\sqrt{2j+1}}\sqrt{(2j+1)!!},
\end{equation}
and since $ (j+1)^{2} \geq 2j+1 $ for all $ j \in \mathbb{N} $, I know $ \frac{j+1}{\sqrt{2j+1}} \geq 1 $, and thus
\begin{equation}
	(j+1)! \geq \sqrt{(2j+1)!!}.
\end{equation}
Since $ 1! = \sqrt{(2\cdot1-1)!!} $, I know by induction that $ j! \geq \sqrt{(2j-1)!!} $ for all $ j \in \mathbb{N} $. Thus, $ \frac{\sqrt{(2j-1)!!}}{j!} \leq 1 $ for all $ j $, which means I could omit this factor from each term of our sum in Eqn. \ref{eq:SumPig} and still have an expression bounding the error in the weak value approximation.

%


\end{document}